\documentclass[11pt,amsmath,amssymb,nofootinbib]{revtex4-2} 
\newcommand{\commentout}[1]{}

\usepackage{graphicx}  
\usepackage{placeins}

\newcommand{\NewOld}[2]{{#1}}
\newcommand{\Add}[1]{{{#1}}}
\newcommand{\st}[1]{}

\newcommand{\half}{\frac{1}{2}}

\newcommand{\SpeciesFlux}{\mathcal{F}}
\newcommand{\ReversibleStress}{\mathcal{R}}
\newcommand{\WhiteNoiseMass}{\mathcal{Z}}
\newcommand{\WhiteNoiseMomentum}{\mathcal{W}}

\newcommand{\InitRad}{{R_0}}
\newcommand{\tPinch}{{t_\mathrm{p}}}

\newcommand{\PinchTime}{{t_\mathrm{p}}}
\newcommand{\OhNum}{{\mathrm{Oh}}}
\newcommand{\ScNum}{{\mathrm{Sc}}}
\newcommand{\WeNum}{{\mathrm{We}^\star}}

\newcommand{\ViscousTensor}{\boldsymbol{\tau}}

\newcommand{\pderiv}[2]{\frac{\partial #1}{\partial #2}}

\begin{document}

\title{Fluctuating Hydrodynamics and the Rayleigh-Plateau Instability}

\author{Bryn Barker}
\email{brynbarker@unc.edu}
\affiliation{Department of Mathematics, The University of North Carolina, Chapel Hill, 120 E Cameron Avenue, Chapel Hill, NC 27599}

\author{John B. Bell} 
\affiliation{Center for Computational Sciences and Engineering, Lawrence Berkeley National Laboratory, 1 Cyclotron Road, Berkeley, CA 94703}

\author{Alejandro L. Garcia}
\affiliation{Department of Physics and Astronomy, San Jose State University, 1 Washington Square, San Jose, CA 95192}

\begin{abstract}
The Rayleigh-Plateau instability occurs when surface tension makes a fluid column become unstable to small perturbations. 
At nanometer scales, thermal fluctuations are comparable to 
interfacial energy densities.  Consequently, at these scales, thermal fluctuations play a significant role in the dynamics of the instability.
These microscopic effects have previously been investigated numerically using particle-based simulations, such as molecular dynamics, and stochastic partial differential equation based hydrodynamic models, such as stochastic lubrication theory.
In this paper we present an incompressible fluctuating hydrodynamics model with a diffuse-interface formulation for binary fluid mixtures designed for the study of stochastic interfacial phenomena.
An efficient numerical algorithm is outlined and validated in numerical simulations of stable equilibrium interfaces.
We present results from simulations of the Rayleigh-Plateau instability for long cylinders pinching into droplets for Ohnesorge numbers of Oh = 0.5 and 5.0.
Both stochastic and perturbed deterministic simulations are analyzed and ensemble results show significant differences in the temporal evolution of the minimum radius near pinching.
Short cylinders, with lengths less than their circumference, were also investigated. 
As previously observed in molecular dynamics simulations, we find that thermal fluctuations cause these to pinch in cases where a perturbed cylinder would be stable deterministically.  Finally we show that the fluctuating hydrodynamics model can be applied to study a broader range of surface-tension driven phenomena.
\end{abstract}

\date{today}

\maketitle

\section*{Introduction}

Liquid jets forming sprays are ubiquitous in Nature and in industrial processes, a familiar example being a stream of water that breaks up into droplets.
The 19th century experiments of Plateau, Beer, and others showed that a long cylinder (length $L$, initial radius $\InitRad$) of fluid (density $\rho$) was unstable to variations that reduced its surface area \cite{Eggers_2008,MontaneroReview2020}.
Plateau predicted that perturbations are unstable for wavelengths $\lambda \geq 2 \pi \InitRad$
and Rayleigh derived that, in the inviscid limit, the fastest growing wavelength is $\lambda_\mathrm{p} \approx 9.01 \InitRad$.
In the Stokes limit (negligible inertia) Tomotika \cite{tomotika1935instability} showed that $\lambda_\mathrm{p} \approx 11.16 \InitRad$ for a fluid cylinder immersed in a similar fluid of equal viscosity.

There are several dimensionless numbers that characterize the dynamics of the Rayleigh-Plateau instability. The Ohnesorge number, $\OhNum = \eta/\sqrt{\rho\InitRad\gamma}$, compares viscous forces to inertial and surface tension forces, characterizing the relative importance of shear viscosity, $\eta$, to surface tension, $\gamma$.
Other dimensionless quantities include the Weber, Bond, and Schmidt numbers that capture the effects of fluid velocity, gravity, and diffusion.

At microscopic scales ($\InitRad \lesssim 10$~nm), thermal fluctuations become important since the variance of velocity fluctuations in a volume $V$ goes as $k_B T / \rho V$ where $T$ is temperature and $k_B$ is the Boltzmann constant. 
For interfacial flows this effect can be characterized by a stochastic Weber number $\mathrm{We}^\star = k_B T / \gamma \InitRad^2$ 
based on a thermal velocity $v^\star = \sqrt{k_B T/\rho \InitRad^3}$. 
Thermal energy becomes comparable to interfacial energy at length scale $\ell^\star = \sqrt{k_B T / \gamma}$
so $\mathrm{We}^\star = (\ell^\star/\InitRad)^2$.

In molecular dynamics (MD) simulations by Moseler and Landman ($\mathrm{We}^\star \approx 0.04$)  the fluid cylinder formed a double-cone (or hourglass) shape as it pinched, in contrast to the macroscopic predictions of an extended, thin liquid thread \cite{Nanojets_2000}.
Furthermore, their numerical solutions of a stochastic lubrication equation
were in qualitative agreement with these MD results.
For $\OhNum \gg 1$, Eggers showed that near the pinching time, $\tPinch$, the minimum cylinder radius goes as $(\tPinch - t)^\alpha$ with $\alpha \approx 0.4$ when thermal fluctuations are significant and $\alpha = 1$ when they are negligible \cite{Eggers_02}.
There have been other studies using MD simulations \cite{RP_MD_2006,RP_MD_2014}, Lattice Boltzmann \cite{RayPlatLatBoltz2020}, 
and Dissipative Particle Dynamics (DPD) simulations \cite{RP_DPD_2008,RP_DPD_2021}.
Recently, the group at Warwick has extensively analyzed both the Rayleigh-Plateau instability\cite{RayPlatFHD,Nanothreads_2020,ThreadPinch_2021}
and related thin film phenomena~\cite{ThinFilmMecke2001,ThinFilmGrun2006,CapWaveThinFilm_2021,ThinFilmFluct_2021,CapWaveFluct2023}
with stochastic lubrication theory and MD simulations.

In this paper we use a fluctuating hydrodynamic (FHD) model for numerical simulations of the Rayleigh-Plateau instability. 
The theoretical foundation of our model is the same as that of stochastic lubrication theory, namely the stochastic Navier-Stokes equations introduced by Landau and Lifshitz \cite{Landau_59,Zarate_07}.
Since our multiphase FHD model \cite{MultiphaseFHD2014,RTIL,Gallo2022} does not use the lubrication approximation it has broader applicability, including modeling the instability past the initial pinching time and for a wider range of geometries and initial conditions. 
The next section outlines the model, followed by a description of the algorithm and its validation. 
Numerical results for the Rayleigh-Plateau instability are then presented for a variety of scenarios. We conclude with a summary of the current work and potential future studies.

\section*{Fluctuating hydrodynamic theory}

We consider a binary mixture of similar species (molecule mass $m$) at constant density and temperature.
We model the specific free energy density of the mixture using the Cahn-Hilliard formalism with regular solution theory \cite{CahnHilliard:1958} and write,
\begin{align}
\frac{\mathcal{G}}{\rho k_B T} =
c \ln c + (1-c) \ln (1-c) 
+ \chi c (1-c) 
+  \kappa |\nabla c |^2 
\label{eq:CHfree} 
\end{align}
where $c$ is the mass fraction of one of the species.
The interaction coefficient is $\chi = 2 T_c/T$ where $T_c$ is the critical temperature. For $\chi > 2$ the mixture phase separates into concentrations $c_{e,1}$ and $c_{e,2}$ given by
\begin{align}
\ln \left( \frac{c_e}{1-c_e}\right) = \chi (2 c_e - 1)    
\end{align}
The surface energy coefficient is $\kappa$ and the surface tension is
\begin{align}
    \gamma = n k_B T \sqrt{2 \chi \kappa} ~\sigma_r
    \label{eq:SurfaceTension}
\end{align}
where $n = \rho/m$ is the number density and
\begin{align}
    \sigma_r = \int_{c_{e,1}}^{c_{e,2}} dc 
    \left[  \frac{2c}{\chi} \ln \frac{c}{c_e} + \frac{2(1-c)}{\chi} \ln \frac{1-c}{1-c_e} - 2 (c - c_e)^2 \right]^{1/2}
\end{align}
The expected surface interface thickness is
\begin{align}
    \ell_\mathrm{s} = \sqrt{2}\ell_\mathrm{c} \left(-1-\frac{2\log{4c_e(1-c_e)}}{\chi(1-2c_e)^2}\right)^{-1/2},
    \label{eq:InterfaceThickness}
\end{align}
where $\ell_\mathrm{c} = \sqrt{2\kappa / \chi}$ is a characteristic length scale for the interface. The characteristic length scale for capillary wave fluctuations is $\ell^\star = \sqrt{ k_B T / \gamma} = (2 \chi \kappa n^2 \sigma_r^2)^{-1/4}$.

For systems in which the characteristic fluid velocity is asymptotically small relative to the sound speed, we can obtain the low Mach number equations from the fully compressible equations by asymptotic analysis \cite{Klainerman:1982,Majda:1985}. 
For constant density this give the incompressible flow equations
\begin{align}
( \rho c)_t + \nabla \cdot(\rho u c) =& \nabla \cdot {\SpeciesFlux}   \nonumber \\
( \rho u)_t + \nabla \cdot(\rho u u) + \nabla \pi =& \nabla \cdot {\ViscousTensor}  + \nabla \cdot \ReversibleStress \nonumber \\
\nabla \cdot u =& 0
\label{eq:low_mach_eqs}
\end{align}
where $u$ is the fluid velocity and $\pi$ is a perturbational pressure.
Here, $\SpeciesFlux$, $\ViscousTensor$, and $\ReversibleStress$ are the species flux, viscous stress tensor, and the interfacial reversible stress, respectively.

In fluctuating hydrodynamics the dissipative fluxes are written as the sum of deterministic and stochastic terms. 
The species flux is $\SpeciesFlux = \overline{\SpeciesFlux} + \widetilde{\SpeciesFlux}$ where the deterministic flux is
\begin{align}
\overline{\SpeciesFlux} = \rho D \left ( \nabla c - 2 \chi  c (1-c) \nabla c +  2c(1-c) \kappa \nabla \nabla^2 c \right )
\end{align}
and $D$ is the diffusion coefficient. 
The stochastic flux is
\begin{align} 
\widetilde{\SpeciesFlux} = \sqrt{2 \rho m D c (1-c)}  ~\WhiteNoiseMass
\end{align}
where $\WhiteNoiseMass(\mathbf{r},t)$ is a standard Gaussian white noise vector with uncorrelated components,
\begin{align}
\langle {\WhiteNoiseMass}_{i}(\mathbf{r},t)
{\WhiteNoiseMass}_{j}(\mathbf{r}',t') \rangle =
\delta_{i,j} \delta(\mathbf{r}-\mathbf{r}') \delta(t-t')
\end{align}

The viscous stress tensor is given by
$\ViscousTensor =  \overline{\ViscousTensor} +  \widetilde{\ViscousTensor}$
where the deterministic component is
\begin{align}
\overline{\ViscousTensor} = \eta [\nabla u + (\nabla u)^T]
\end{align}
Here, bulk viscosity is neglected because it does not appear in the incompressible equations.
The stochastic contribution to the viscous stress tensor is modeled as,
\begin{align}
\widetilde\ViscousTensor = \sqrt{\eta k_B T}({\WhiteNoiseMomentum} + {\WhiteNoiseMomentum}^T),
\end{align}
where 
${\WhiteNoiseMomentum}(\mathbf{r},t)$ is a standard Gaussian white noise tensor
with uncorrelated components.
Finally, the interfacial reversible stress is
\begin{align}
\ReversibleStress = n k_B T \kappa \left [\frac{1}{2} |\nabla c|^2 \mathbb{I} - \nabla c \otimes \nabla c \right].
\label{eq:reversible_stress}
\end{align}
Note that since $\ReversibleStress$ is a non-dissipative flux there is no corresponding stochastic flux.

\section*{FHD algorithm and its validation}

The system of equations [\ref{eq:low_mach_eqs}] is discretized using a structured-grid finite-volume approach with cell-averaged concentrations and face-averaged (staggered) velocities. The overall algorithm is based on methods introduced in~\cite{donev2014low, Donev_10, RTIL}.  The algorithm uses an explicit discretization of concentration coupled to a semi-implicit discretization of velocity using a predictor-corrector scheme for second-order temporal accuracy.

The numerical method uses standard spatial discretization approaches.  Details appear in \Add{Section A of} the Supporting Information, which is at the end of the paper. The basic time step algorithm consists of four steps:

\noindent
{\bf{Step 1:}} Compute the predicted velocity , $u^{*,n+1}$ and perturbational $\pi^{*,n+\frac{1}{2}}$, by solving the Stokes system
\begin{align}
    \frac{\rho u^{*,n+1}-\rho u^n}{\Delta t} &+ \nabla \pi^{*,n+\frac{1}{2}} = -\nabla \cdot (\rho u u^T)^n \\
    &+\frac{1}{2} \left(\nabla \cdot \overline{\ViscousTensor}^n + \nabla \cdot \overline{\ViscousTensor}^{*,n+1} \right)  \nonumber \\
&+ \nabla \cdot \widetilde{\ViscousTensor}^n + \nabla \cdot \ReversibleStress^n \nonumber \nonumber \\
\nabla \cdot u^{*,n+1} &= 0
\end{align}
\noindent
{\bf{Step 2:}}  Predict concentration at time $n+1/2$ using
\begin{align}
    \rho c^{*,n+\half} &= \rho c^n- \frac{\Delta t}{2} \nabla \cdot \left [ \rho c^n \left(\frac{ u^n + u^{*,n+1}}{2}\right) \right ] \\
    &+\frac{\Delta t}{2} \left(  \nabla \cdot \overline{\SpeciesFlux}^n + \nabla \cdot \widetilde{\SpeciesFlux}^n \right) \nonumber
\end{align}
{\bf{Step 3:}}  Compute concentration at time $n+1$ using
\begin{align}
    \rho c^{n+1} &= \rho c^n- {\Delta t} \nabla \cdot \left [ \rho c^{*,n+\half}
    \left(\frac{ u^n + u^{*,n+1}}{2}\right) \right ] \\
    &+{\Delta t} \left(  \nabla \cdot 
    \overline{\SpeciesFlux}^{*,n+\half}+ \nabla \cdot \widetilde{\SpeciesFlux}^{*,n+\half} \right) \nonumber
\end{align}
\noindent
{\bf{Step 4:}}  Compute the corrected velocity $u^{n+1}$ and perturbational pressure $\pi^{n+\frac{1}{2}}$, by solving the Stokes system
\begin{align}
    \frac{\rho u^{n+1}-\rho u^n}{\Delta t} &+ \nabla \pi^{*,n+\frac{1}{2}} = -\frac{ \left( \nabla \cdot (\rho u u)^n 
    + \nabla \cdot (\rho u u)^{*,n+1} \right)}{2} \nonumber \\
    &+\frac{1}{2} \left(\nabla \cdot \overline{\ViscousTensor}^n 
    + \nabla \cdot \overline{\ViscousTensor}^{n+1} \right)    \\
&+ \nabla \cdot \widetilde{\ViscousTensor}^n + \nabla \cdot \ReversibleStress^{*,n+\half} \nonumber\\
\nabla &\cdot u^{n+1} = 0
\end{align}
In both Steps 1 and 4, the discretized Stokes system is solved by a generalized minimal residual (GMRES) method with a multigrid preconditioner, see \cite{cai:2014}.
The explicit treatment of the concentration equation introduces a stability limitation on the time step of
\begin{align}
D \left( \frac{12}{\Delta x^2} + \frac{72 \kappa}{\Delta x^4} \right) \Delta t \leq 1
\label{eq:StableDt}
\end{align}
where $\Delta x$ is the mesh spacing.

Unless otherwise specified, the physical parameters used in all simulations are: 
mass density, $\rho = 1.4 ~\mathrm{g/cm}^3$, 
molecular mass, $m = 6.0\times 10^{-23}$~g, 
Boltzmann constant, $k_B = 1.38\times 10^{-16}$~erg/K,
temperature $T = 84$K, 
interaction parameter $\chi = 3.571$, 
surface energy coefficient $\kappa = 2.7 \times 10^{-14} \mathrm{cm}^2$.
These conditions are based on the model for liquid argon given in \cite{RayPlatFHD}, modified to increase both the interfacial tension and the interface thickness by a factor of 2.
For these values the equilibrium concentrations are $c_{1,e} = 0.035$ and $c_{2,e} = 0.965$.
The surface tension is $\gamma = 28.35$~dyne/cm
and $\ell^\star = 0.2$~nm;
for $\InitRad = 6$~nm the stochastic Weber number $\mathrm{We}^\star \approx 10^{-3}$.

We considered two values for shear viscosity, $\eta = 2.46\times 10^{-3}$ and $2.46\times 10^{-2}~\mathrm{g}/\mathrm{cm\,s}$, for which  
the Ohnesorge numbers are $\OhNum = 0.50$ and $5.0$, respectively. 
In general the diffusion coefficient was $D = \eta/\rho\,\mathrm{Sc}$ with a Schmidt number of $\mathrm{Sc} = 35.1$; the exception being a single run with $\mathrm{Sc} = 351$ (see Fig.~\ref{fig:HighViscMinRadius}).

In general, the simulations used periodic boundary conditions and cubic cells with mesh spacing 
$\Delta x = 1.0$~nm.  
With these parameters each simulation cell represents roughly 23 fluid molecules.
The time step was either $\Delta t = 0.4$~ps or 0.04~ps depending on the
value of $D$, which corresponds to approximately one quarter of the maximum stable time step (see Eq.~\ref{eq:StableDt}). 

\begin{figure}
  \centering
  \includegraphics[width=.95\textwidth]{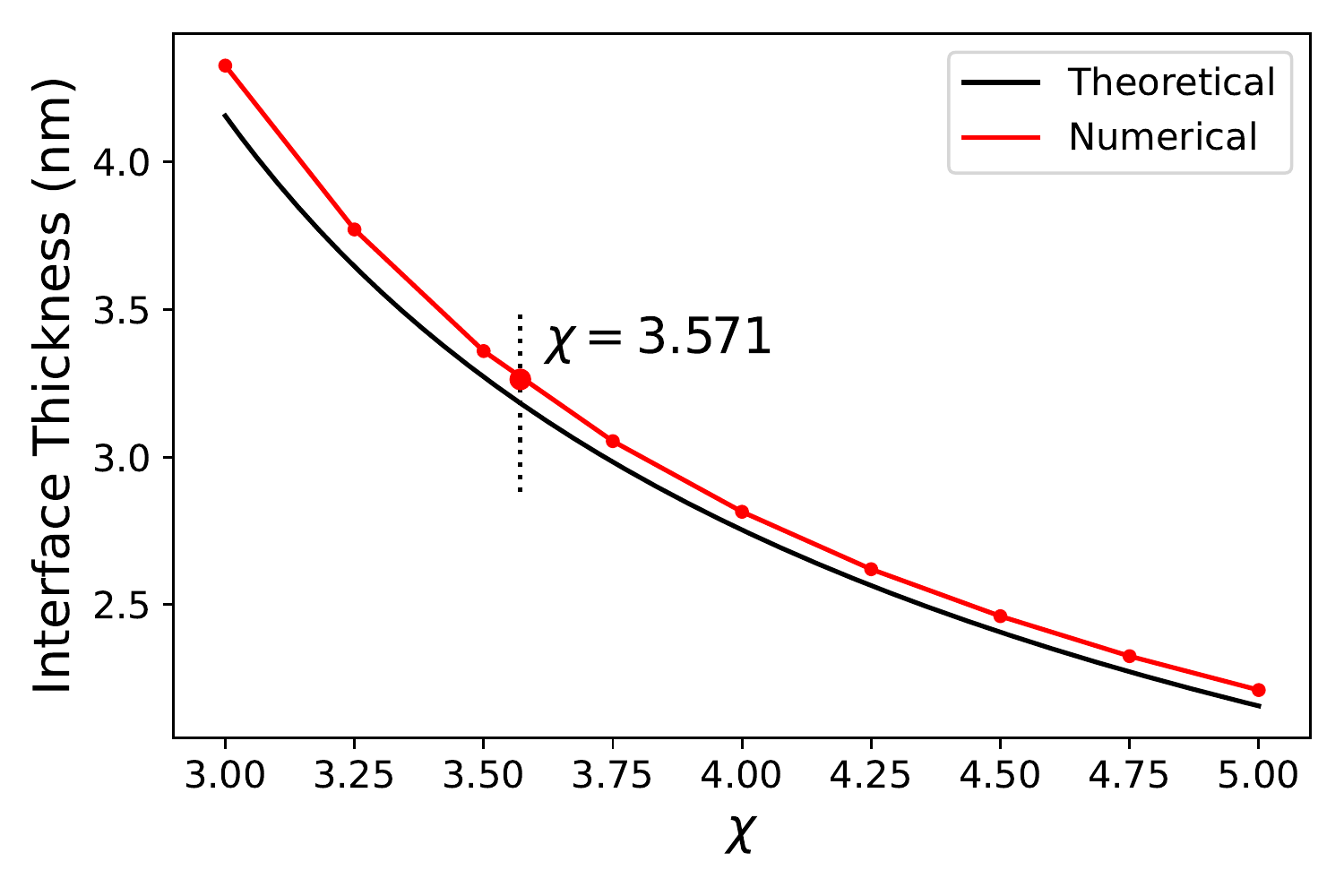}
  \caption{Interface thickness, $\ell_\mathrm{s}$, in nm versus interaction coefficient $\chi$ as measured (markers) and as given by Eq.~\ref{eq:InterfaceThickness} (solid curve); large marker indicates the value of $\chi$ used in the Rayleigh-Plateau simulations.
  Note that $\ell_\mathrm{s} \rightarrow \infty$ as $\chi \rightarrow 2$.
  }
  \label{fig:ThicknessGraph}
\end{figure}

A variety of equilibrium systems were simulated to validate the algorithm.
First, the interface thickness was measured in the simulations
of a flat slab in a
quasi-2D system ($96 \times 12 \times 1$ cells)\footnote{See \Add{Section B of} the Supporting Information for details of how interface thickness is measured \Add{and how the estimated thickness depends on resolution}.}; 
Fig.~\ref{fig:ThicknessGraph} shows that good agreement with Eq.~\ref{eq:InterfaceThickness} is found.
\Add{The systematic shift in the predictions is a result of numerical error, which decreases with mesh spacing.}
Note that the surface interface thickness $\ell_\mathrm{s} \approx 3 \Delta x \approx 15 \ell^\star$.

\begin{figure}
  \centering
  \includegraphics[width=.95\textwidth]{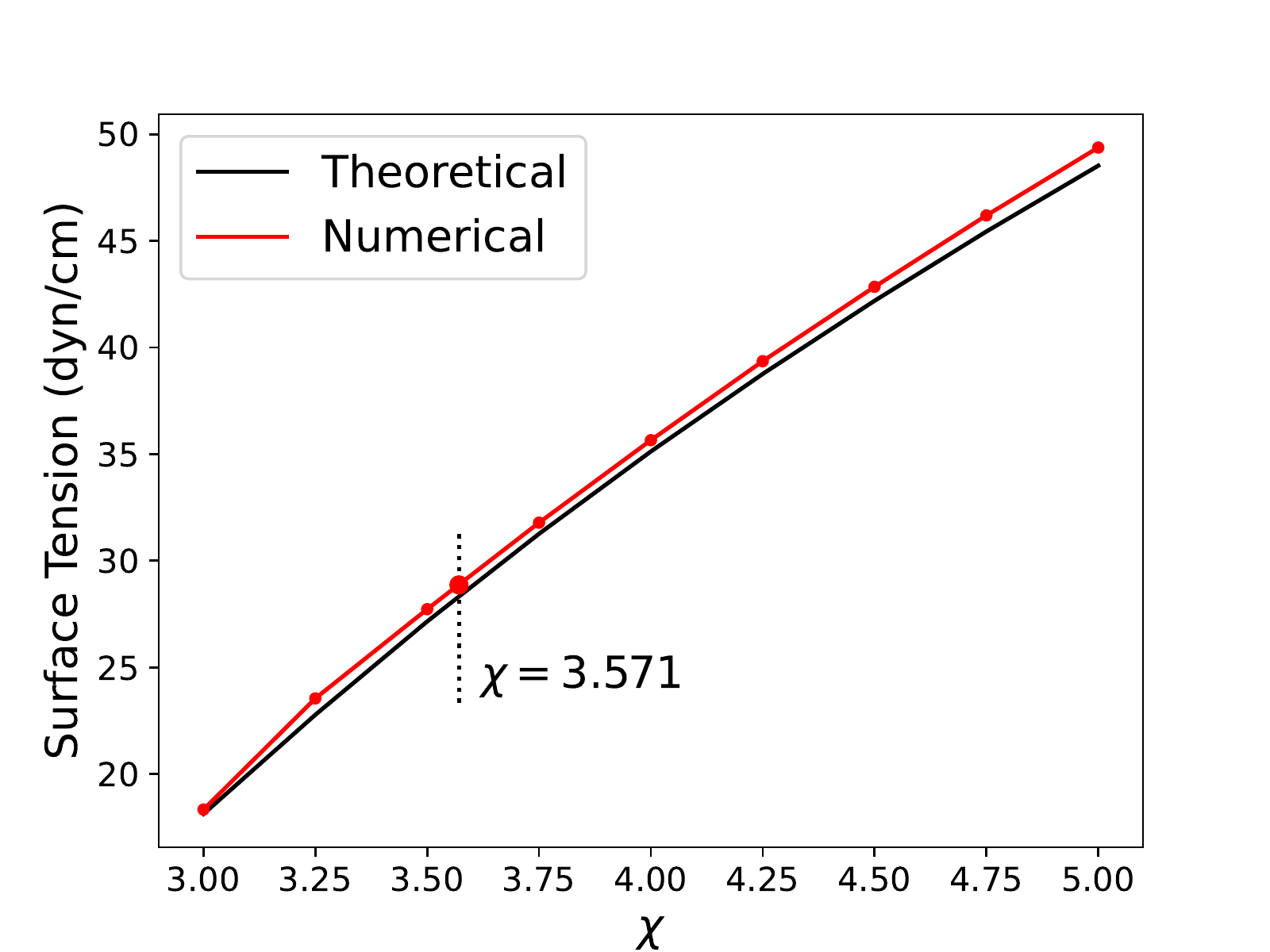}
  \caption{Surface tension, $\gamma$, in dyne/cm versus interaction coefficient $\chi$ as measured (markers) and as predicted by Eq.~\ref{eq:SurfaceTension} (solid curve); large marker indicates value of $\chi$ used in the Rayleigh-Plateau simulations.
  Note that $\gamma \rightarrow 0$ as $\chi \rightarrow 2$.}
  \label{fig:SurfaceTensionGraph}
\end{figure}

Next, the Laplace pressure, $\delta p$, was measured
in a similar quasi-2D system ($96 \times 96 \times 1$ cells) with concentration $c_{1,e}$ within a disk of radius $\InitRad = 6.0$~nm and concentration $c_{2,e}$ elsewhere.\footnote{The validation tests of interface thickness and surface tension were performed without the stochastic noises (i.e., by setting $\widetilde{\SpeciesFlux} = \widetilde{\ViscousTensor} = 0)$.}
Figure~\ref{fig:SurfaceTensionGraph} shows that the surface tension, computed using $\gamma = \InitRad \delta p$, is in good agreement with the expected value given by Eq.~\ref{eq:SurfaceTension}.\footnote{\Add{Details of how the surface tension was computed are discussed in Section C of the Supporting Information.}}
As part of this validation we also considered disks of different radii and alternative parameters that resulted in a thinner interface.  In all of these additional cases, the methodology continued to show excellent agreement with theory.

\begin{figure}
  \centering
  \includegraphics[width=.95\textwidth]{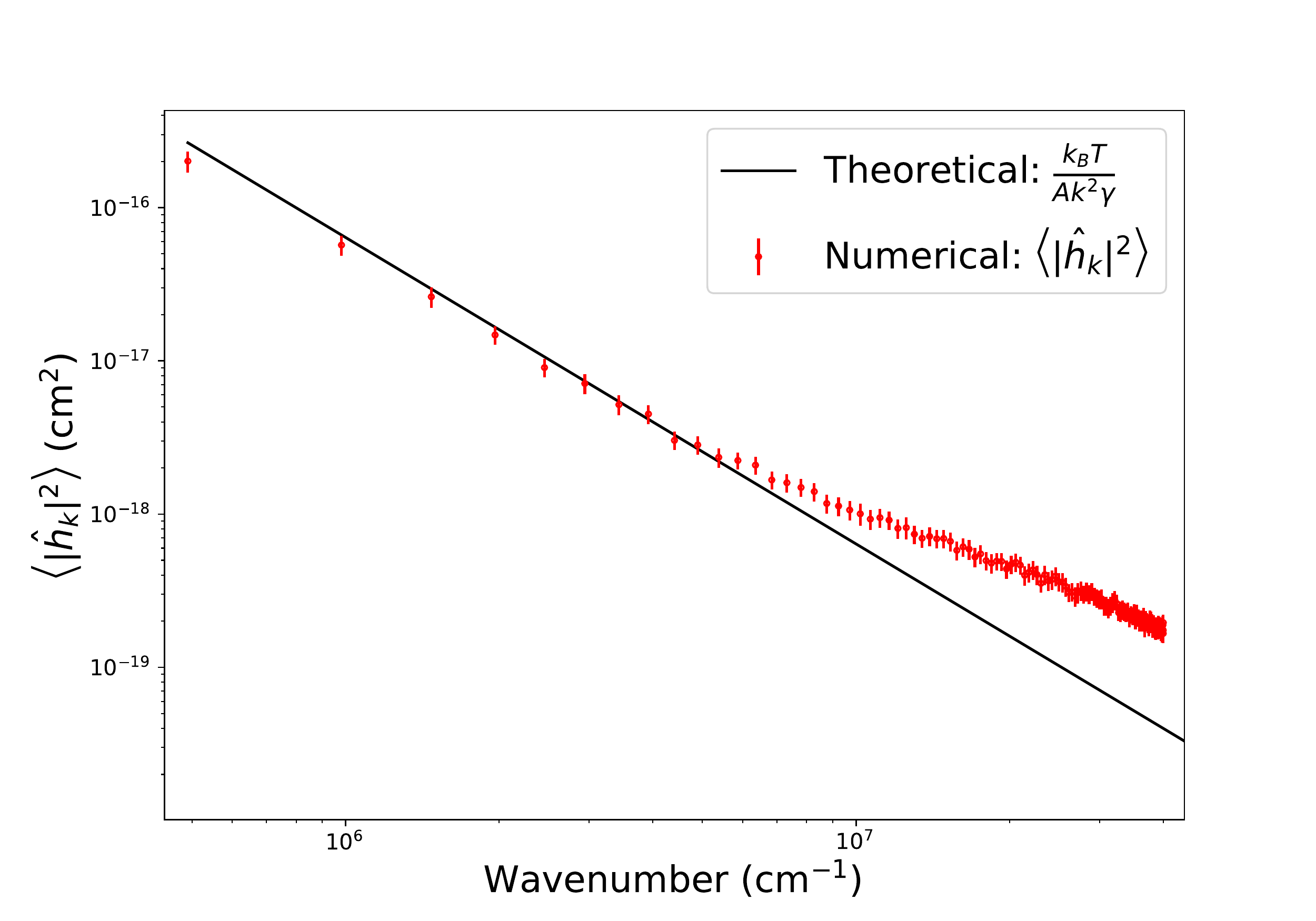}
  \caption{Capillary wave spectrum (in $\mathrm{cm}^2)$ versus wavenumber (in $\mathrm{cm}^{-1}$) as measured in the FHD simulation (\NewOld{points}{dotted line}) and as predicted by Eq.~\ref{eq:CatPaw} (solid line).
  \Add{The error bars represent $\pm$3 standard deviations.}
  The wavenumber is corrected to account for the discrete Laplacian (see \cite{MultiphaseFHD2014})}.
  \label{fig:CatPawGraph}
\end{figure}

As a final validation test the capillary wave spectrum \cite{CapWaveTheory1913,CapWaveExperiment1968,CapWaveExperimentScience2004} at thermal equilibrium was measured in a quasi-2D system ($256 \times 64 \times 1$ cells) with a flat slab of concentration $c_{1,e}$ and concentration $c_{2,e}$ elsewhere. As in \cite{MultiphaseFHD2014}, the deviations in height from a flat interface, $h(\mathbf{r},t)$, were measured and Fourier transformed to obtain $\hat{h}(\mathbf{k},t)$. The temporal averaged spectrum, shown in Fig.~\ref{fig:CatPawGraph}, is in good agreement with the predicted result,
\begin{align}
    \langle |\hat{h}(\mathbf{k}) |^2 \rangle = \frac{k_B T}{A \gamma k^2}
    \label{eq:CatPaw}
\end{align}
where $A$ is the surface area of the interface (here the simulation was two-dimensional with a cross-section of 5~nm, which sets the magnitude of the noise).
The deviation from theory at large $k$ is attributed to a wave vector-dependent surface tension. It is also observed in molecular dynamics \cite{CapWaveMD2011} and the deviation is sensitive to how the interface position is defined \cite{Tarazona_2012}.

\section*{Rayleigh-Plateau instability}

The Rayleigh-Plateau instability in nanoscale systems was simulated and compared with earlier molecular dynamics and stochastic lubrication calculations.
Each simulation was initialized by starting with concentration $c_{e,1}$ inside a 2D disk of radius $\InitRad = 6.0$~nm and concentration $c_{e,2}$ elsewhere.
The system was then evolved deterministically until the interface had equilibrated.  This initial slice was replicated to create a uniform cylinder of length $L$. 
Unless otherwise stated, the physical and numerical parameters are those used in the validation runs (see previous section).
The characteristic time scale for capillary waves is $\tau_0 = \sqrt{\rho \InitRad^3/\gamma} \approx 0.1$~ns.
From linear stability theory the growth rate for the fastest growing wavenumber (Rayleigh mode) is
$\tau_\mathrm{inv} \approx 3\, \tau_0$ in the inviscid limit;
for uniform viscosity it is $\tau_\mathrm{visc} \approx 28 \,\OhNum\,\tau_0$~\cite{Eggers_2008,stone_brenner_1996}. 

\begin{figure*}
  \centering
  \includegraphics[width=.95\textwidth]{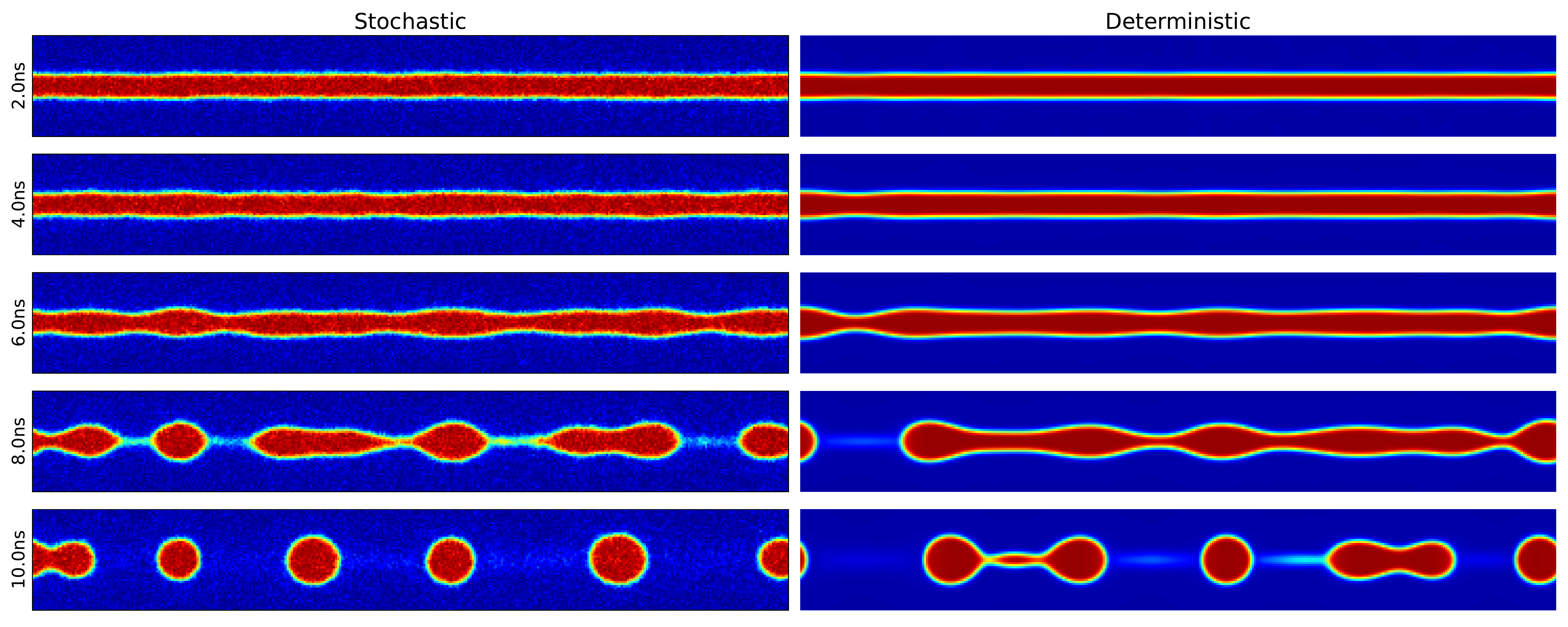}
  \caption{Center-line cross section snapshots from stochastic (left) and deterministic runs at $t = 2$, 4, 6, 8, and 10~ns for $\mathrm{Oh} = 0.5$. Note the double cone shape in the stochastic case at 8~ns and the filament attached to a drop in the deterministic case at 10~ns.}
  \label{fig:RPsnapshotsLong}
\end{figure*}

First we consider ``long'' cylinders with $L = 360.0$~nm so $L \gg \lambda_\mathrm{p} \approx 67$~nm, the fastest growing wavelength.
The domain of this 3D system has a cross section of $48.0$~nm by $48.0$~nm in the $x$ and $y$ directions ($48 \times 48 \times 360$~cells) and  is periodic in all directions.
We ran both a ``stochastic'' and a ``deterministic'' version of the simulation. The former simply uses the FHD algorithm starting from the initial condition described above. 
The deterministic runs start the same way but after a time $t_\mathrm{init} = 0.4$~ns 
the stochastic fluxes are set to zero. 
In both the stochastic and deterministic runs the initial cylinders pinch into droplets but there are qualitative differences, as seen in Figure~\ref{fig:RPsnapshotsLong}.
In the stochastic case the cylinders narrow into a double-cone before pinching while in the deterministic case a filament forms, as seen by comparing stochastic simulation at 8.0~ns and deterministic simulation at 10.0~ns.

\begin{figure}    
  \centering
  \includegraphics[width=.65\textwidth]{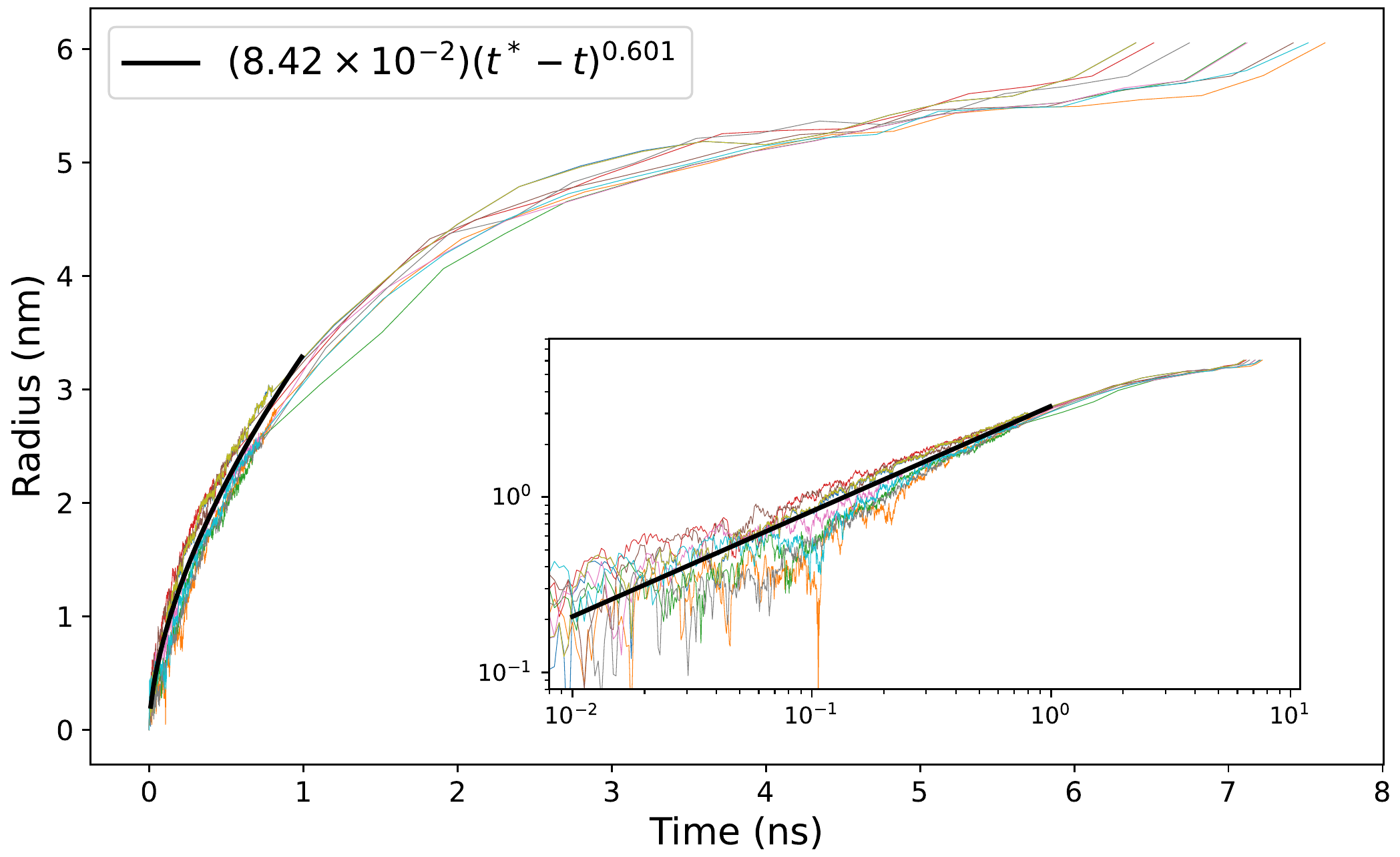} \\
   \includegraphics[width=.65\textwidth]{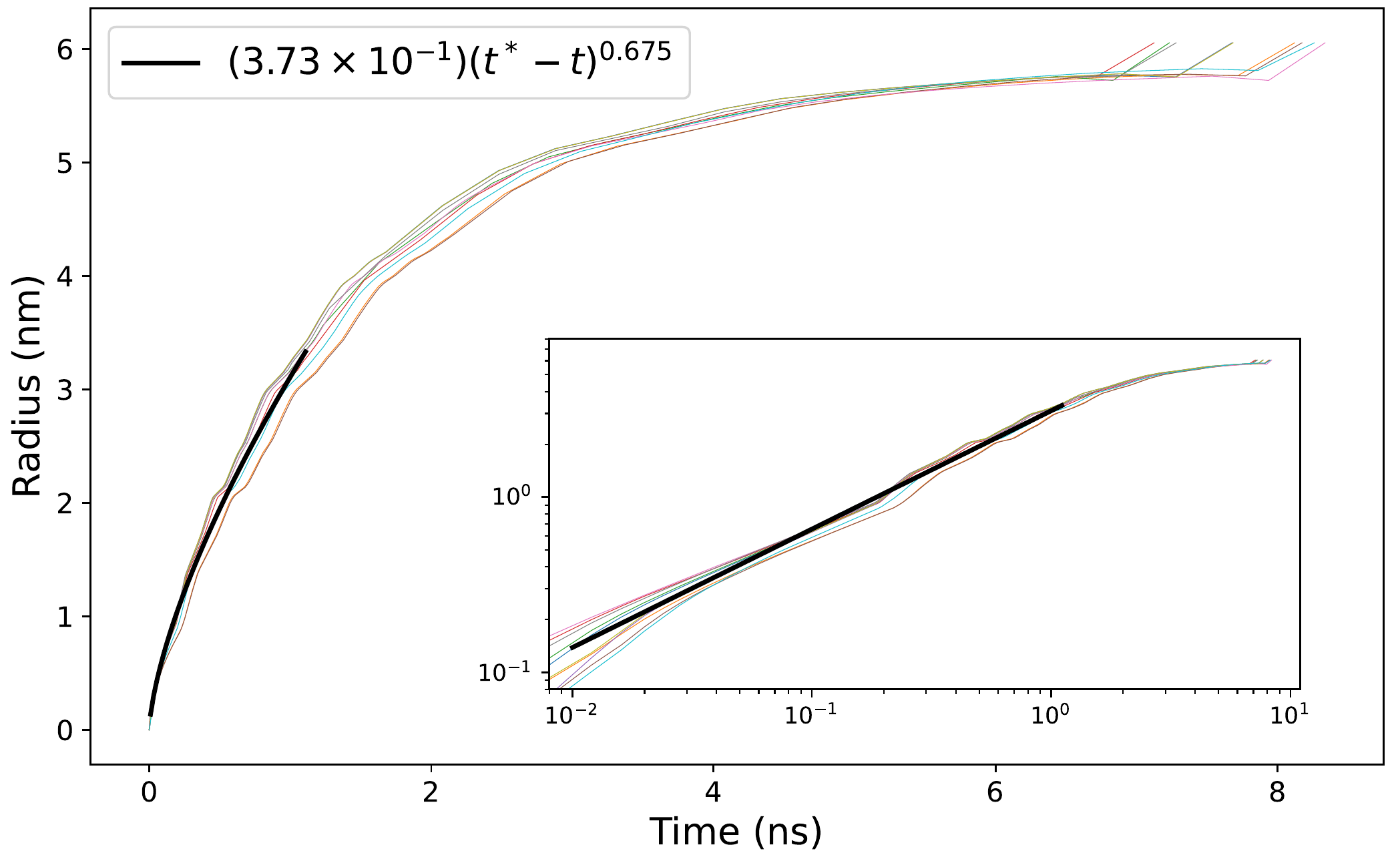} \\
    \includegraphics[width=.65\textwidth]{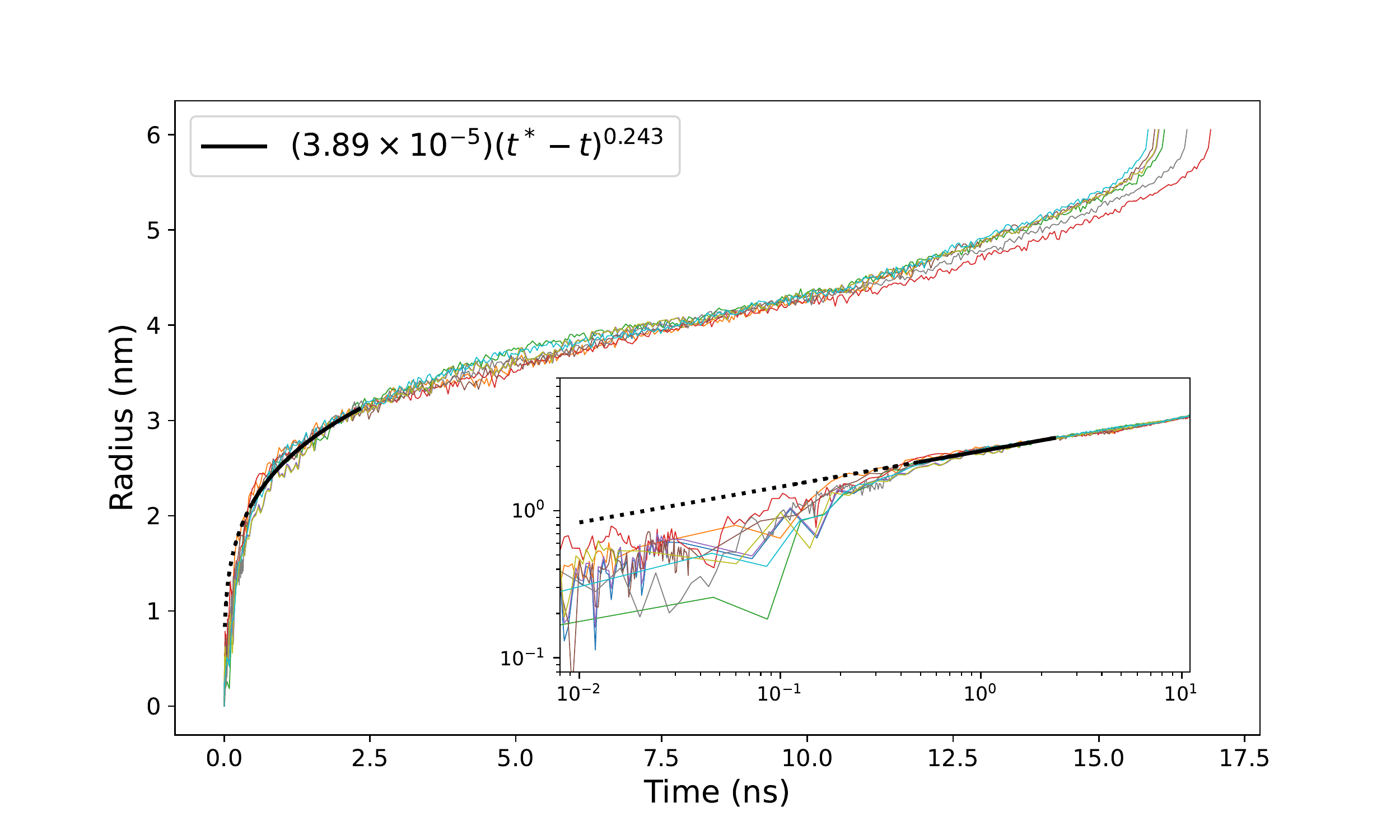}


   
   \caption{
   Minimum radius,  $R_\mathrm{min}(t)$, versus time before pinching for each run (thin lines): (top) stochastic runs, $\mathrm{Oh} = 0.5$; (middle) deterministic runs, $\mathrm{Oh} = 0.5$; (bottom) stochastic runs, $\mathrm{Oh} = 5.0$.
   Average pinch times: (top) 6.93~ns, (middle) 7.73~ns, (bottom) 16.18~ns. Standard deviations: (top) 0.156~ns, (middle) 0.142~ns, (bottom) 0.111~ns.  
   Thick solid lines are power-law fits to the ensemble average; dotted lines are extrapolations.
   Fits give are:
   $\alpha = 0.601$ (top), $\alpha = 0.675$ (middle), $\alpha = 0.243$ (bottom).
   }
  \label{fig:MinRadiusGraph}
\end{figure}   

From the simulation data we calculate the cylinder radius, $R(z,t)$, for each cross section. 
The presence of thermal fluctuations introduces some difficulty in defining the radius so we use a filter to sharpen the numerical interface.\footnote{\Add{The procedure for computing the radius is described in the Section D of the Supporting Information.}}
Figure~\ref{fig:MinRadiusGraph} shows the minimum cylinder radius, $R_\mathrm{min}(t) = \min_z\{R(z,t)\}$, versus time for individual runs in  ensembles of 10 runs. 
For the lower viscosity case (Oh = 0.5) the mean and standard deviation for the pinch time in the stochastic runs were 6.93~ns and 0.156~ns; 
for the perturbed deterministic runs they were 7.73~ns and 0.142~ns
so the fluctuations significantly hasten the breakup of the cylinder.
For the higher viscosity (Oh = 5.0) stochastic runs mean and standard deviation were 16.18~ns and 0.111~ns.

Theory, simulations, and experiments indicated that, in general, $R_\mathrm{min} \sim (\PinchTime - t)^\alpha$ where $\PinchTime$ is the mean pinching time with the 
coefficient $\alpha$ depending on $\OhNum$, $\ScNum$, and $\WeNum$. 
Specifically there are the inertia-dominated (small $\OhNum$), viscosity-dominated (large $\OhNum$, large $\ScNum$), and diffusion-dominated (large $\OhNum$, small $\ScNum$) regimes, which can be either deterministic (small $\WeNum$) or stochastic (large $\WeNum$).
For example, $\alpha = 1$ in the deterministic, viscosity-dominated regime~\cite{RayPlatTwoFluidPRL1999,Eggers_2008}, 
$\alpha = 1/3$ in the deterministic, diffusion-dominated regime~\cite{DiffusionRayPlat2022}, and
$\alpha \approx 0.412$ in the stochastic, viscosity-dominated regime~\cite{Eggers_02}. 
The power-law fits to the simulation data are shown in Fig.~\ref{fig:MinRadiusGraph}
for the $\OhNum = 0.5$ and 5.0 runs with $\ScNum = 35.1$\footnote{See \Add{Section E of the} Supporting Information for details of the fitting procedure.}.
As expected, the $\OhNum = 0.5$ are in an intermediate range between inertia-dominated and diffusion-dominated, similar to the molecular dynamics simulations in \cite{Nanothreads_2020}.
Quantitative comparison was not possible since that work investigated cylinders of liquid in its own vapor while our current results are for two similar incompressible fluids.

\begin{figure}
  \centering
    \includegraphics[width=0.95\textwidth]{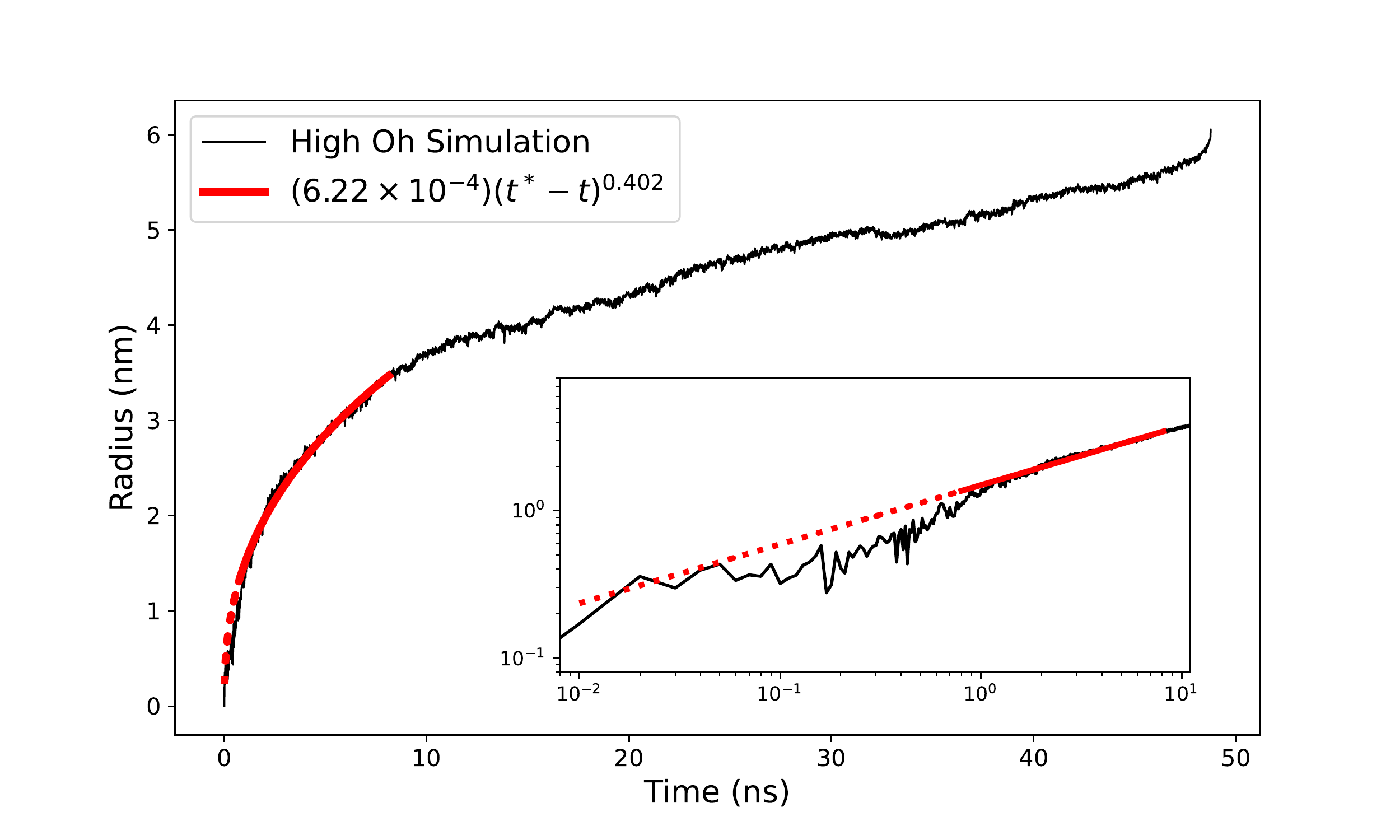}
    \caption{
    Minimum cylinder radius, $R_\mathrm{min}(t)$, versus time time before pinching for a high Schmidt number ($\mathrm{Sc} = 351$), high Ohnesorge number ($\mathrm{Oh} = 5.0$) run.
    Red solid line is power-law fit ($\alpha = 0.402$); dotted line is an extrapolation.
    } 
  \label{fig:HighViscMinRadius}
\end{figure}

A single run was performed using the higher viscosity ($\OhNum = 5.0$) with a Schmidt number of $\mathrm{Sc} = 351$; all other physical parameters were unchanged. 
This run was computationally intensive because the time to pinching was nearly 50~ns.
Figure~\ref{fig:HighViscMinRadius} shows that $R_\mathrm{min}(t)$ for this run is in good agreement with the prediction by Eggers~\cite{Eggers_02} that $R_\mathrm{min} \sim (\PinchTime - t)^{0.412}$ in the stochastic, viscous-dominated regime. 
A comparison with the lower $\mathrm{Sc}$ number simulations illustrates the importance of diffusion on the pinch-off dynamics.

\begin{figure}
  \centering
   \includegraphics[width=0.95\textwidth]{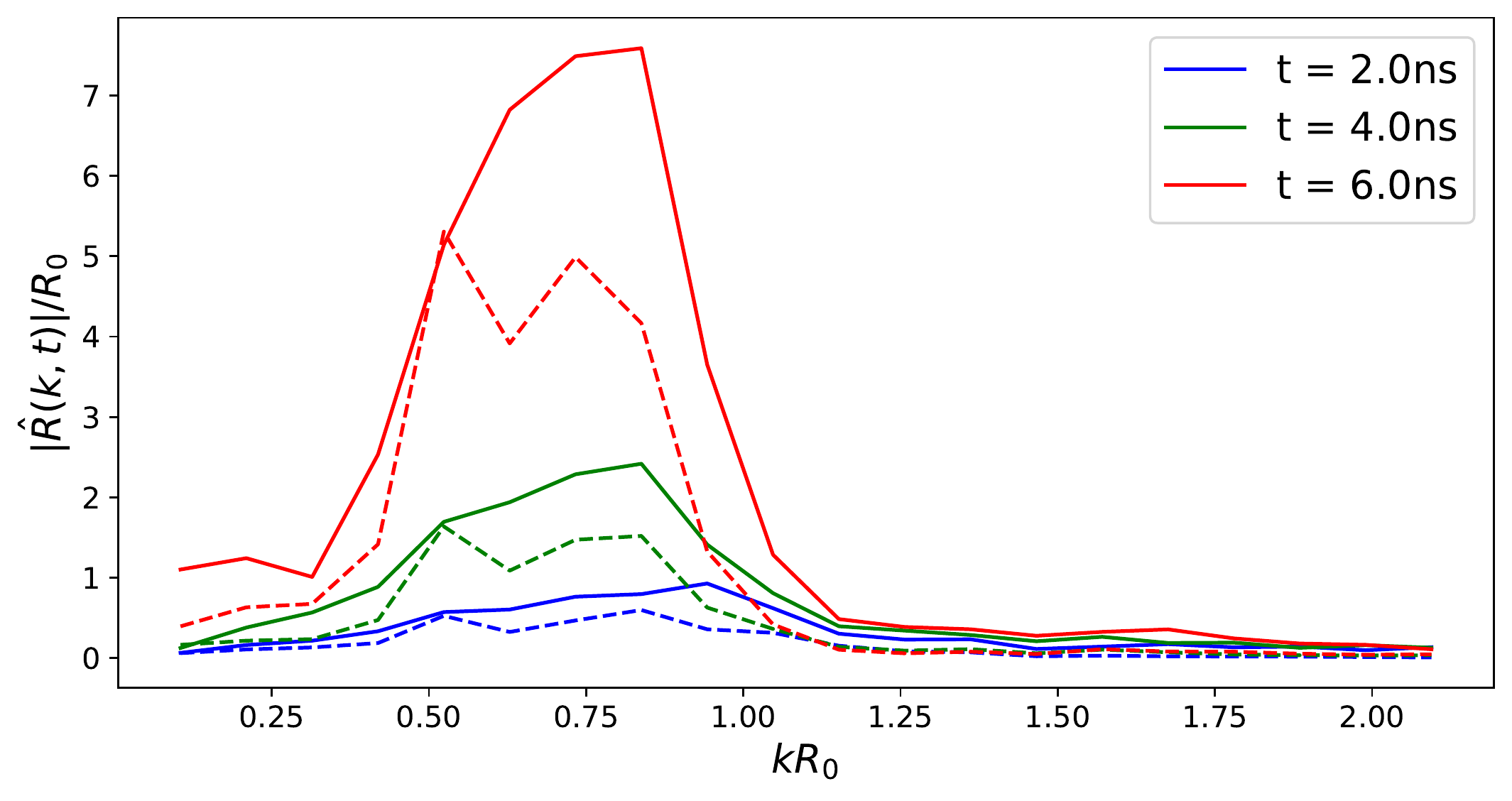}
   \caption{Averaged power spectrum for stochastic (solid lines) and deterministic (dashed lines) simulations with lower viscosity ($\mathrm{Oh} = 0.5$) at $t = 2$, 4, and 6~ns (see legend).}
  \label{fig:FFTgraphs}
\end{figure}

\begin{figure}
  \centering
   \includegraphics[width=0.95\textwidth]{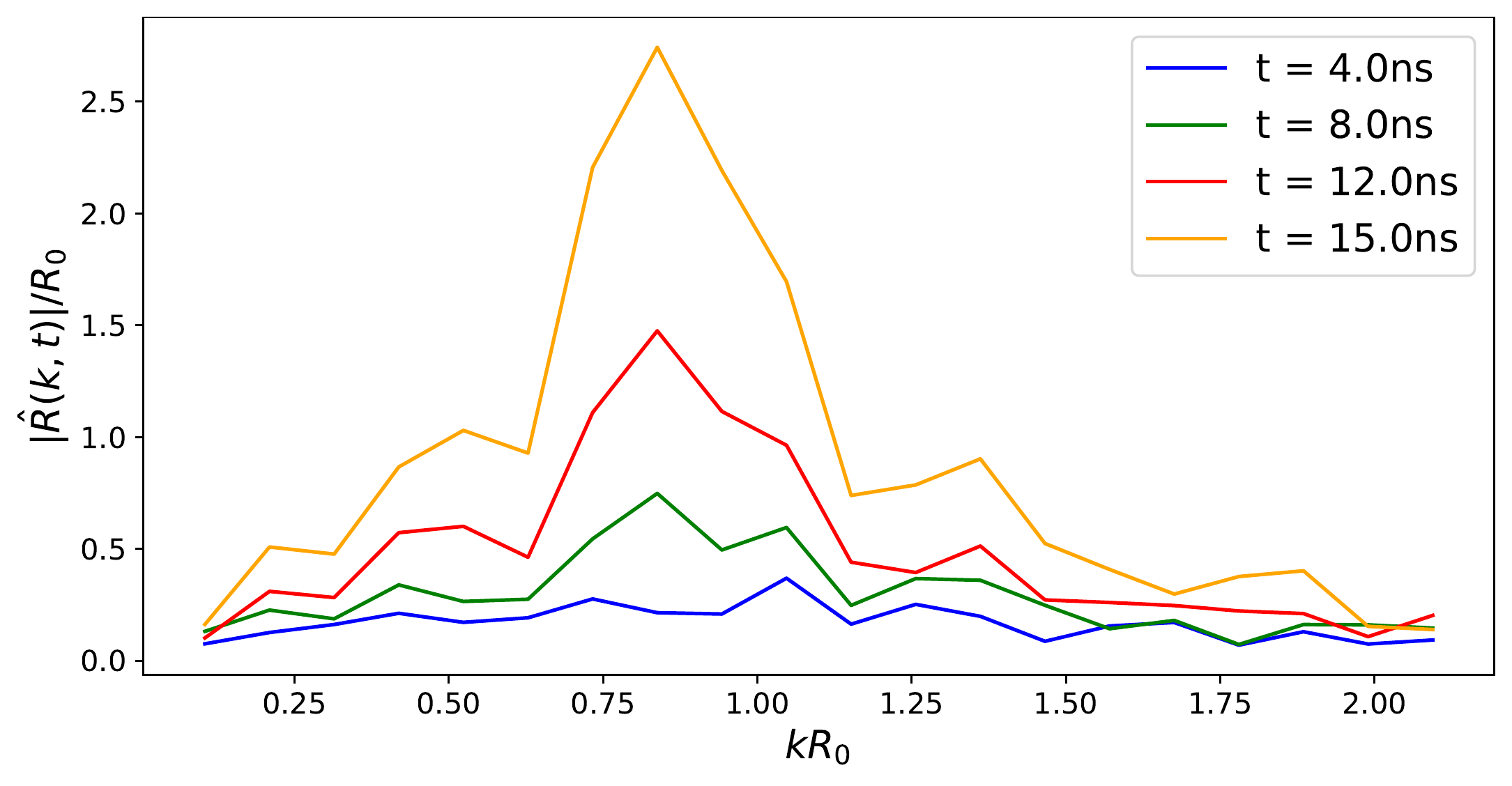}
   \caption{Averaged power spectrum for stochastic simulations with higher viscosity ($\mathrm{Oh} = 5.0$) at $t = 4$, 8, 12 and 15~ns (see legend).}
  \label{fig:HighViscFFTgraphs}
\end{figure}

To quantify the dominant modes of growth leading to rupture,  we 
 Fourier transform $R(z,t)$ to obtain $\hat{R}(k,t)$  for an ensemble of 10 runs, similar to the analysis in Ref.~\cite{RayPlatFHD} (see Supporting Information \Add{Section F}).
Figures~\ref{fig:FFTgraphs} and \ref{fig:HighViscFFTgraphs} show the ensemble averaged spectrum, $|\hat{R}(k,t)|$, versus wave number at various times from simulations using the lower viscosity 
($\eta = 2.46 \times 10^{-3}$~g/cm~s, $\mathrm{Oh} = 0.5$) 
and the higher viscosity ($\eta = 2.46 \times 10^{-2}$~g/cm~s, $\mathrm{Oh} = 5.0$).
In the former case the results from an ensemble of deterministic runs is also shown in Fig.~\ref{fig:FFTgraphs}. 
These results are qualitatively similar to the molecular dynamics measurements and stochastic lubrication prediction in \cite{RayPlatFHD}. Specifically, the spectrum is similar with the instability developing faster in the stochastic case but, again, that work is for a liquid/vapor system. 

\begin{figure}
  \centering
  \includegraphics[width=0.95\textwidth]{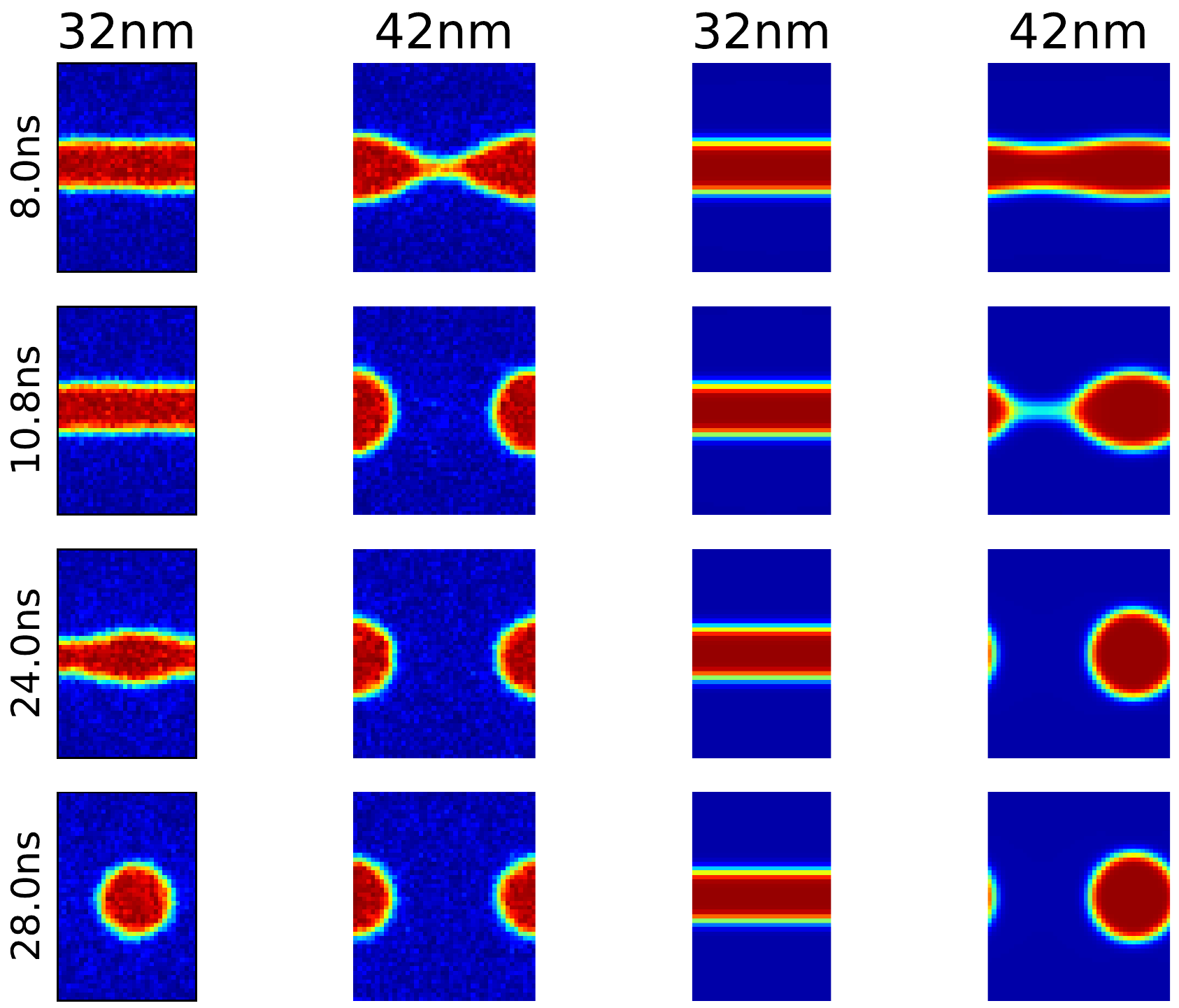}
  \caption{Snapshots of short cylinders with lengths of 32~nm ($L < L_c$) and 42~nm ($L > L_c$). The two columns on the left are stochastic runs and the two on the right are deterministic runs; in all cases $\mathrm{Oh} = 0.5$.}
  \label{fig:StubbySnapshots}
\end{figure}

\begin{figure}[t]
  \centering
  \includegraphics[width=0.95\textwidth]{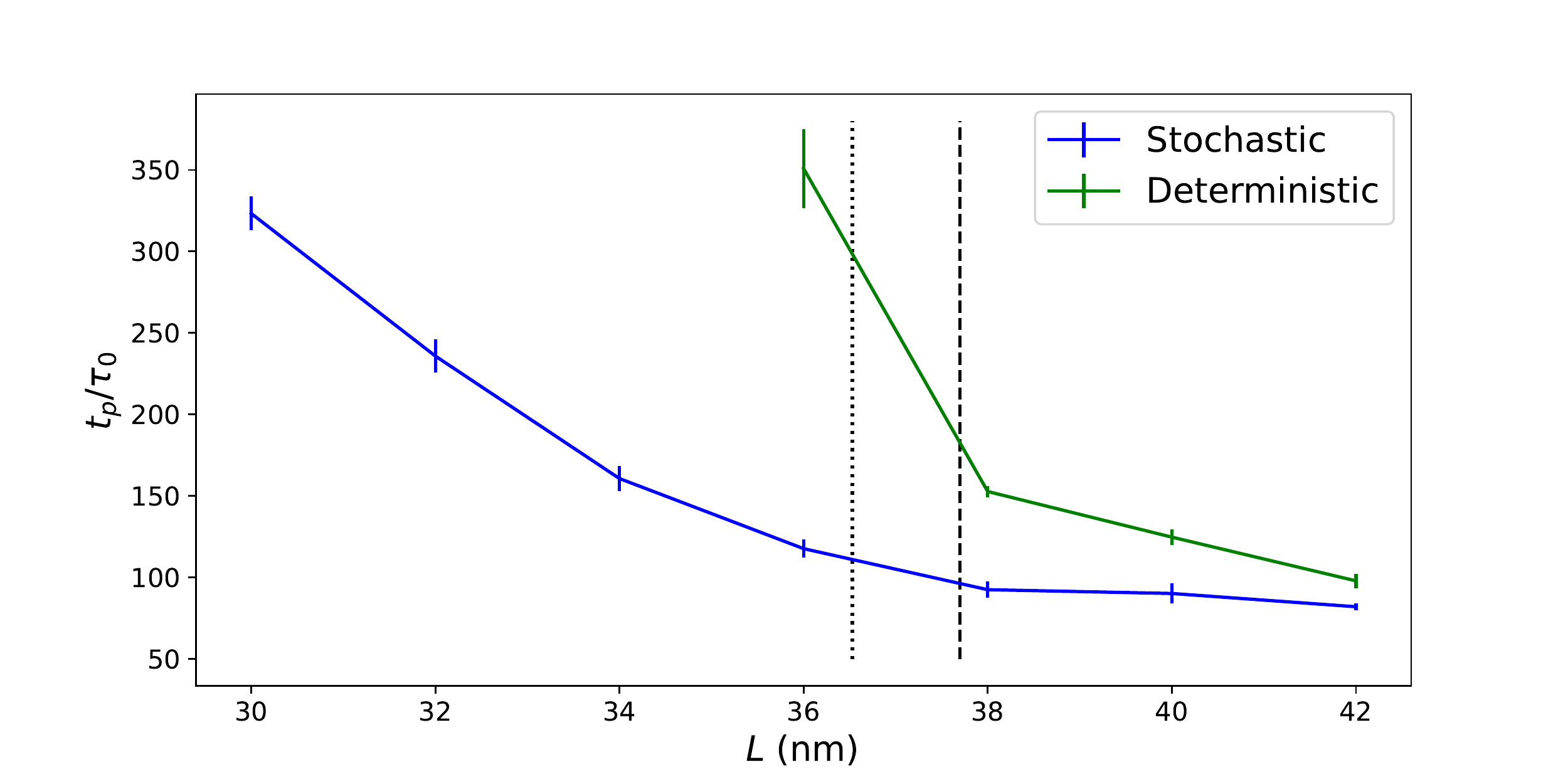}
  \caption{Dimensionless pinch time, $t_\mathrm{p}/\tau_0$, for short cylinders versus $L$ from ensembles of stochastic and deterministic runs. The dashed line marks the critical length (37.7~nm) based on the initial radius and the dotted is the critical length (36.5~nm) based on the minimum radius when noises are turned off in the deterministic runs.}
  \label{fig:PinchTimeGraph}
\end{figure}

We also investigated the breakup of classically stable cylinders, that is, cylinders with length $L < L_c$ where the critical length, $L_c$, equals the circumference.
Figure~\ref{fig:StubbySnapshots} shows snapshots for cylinders of radius $\InitRad = 6.0$~nm ($L_c = 37.7$~nm) for $\mathrm{Oh} = 0.5$.
Note that both stochastic cases pinch to form a droplet while in the deterministic case only the longer cylinder ($L = 42$~nm) forms a droplet. 
Figure \ref{fig:PinchTimeGraph} shows the time to pinching for a range of cylinder lengths; comparable results have been reported for molecular dynamics simulations \cite{RayPlatFHD}.
Interestingly, the deterministic case with $L = 36$~nm is unstable, which suggests that the effective critical length is slightly shorter (deterministic runs with $L \leq 34$~nm did not pinch). 
This is less surprising when we recall that the diffuse interface is relatively thick (see Fig.~\ref{fig:ThicknessGraph}).

\begin{figure}[b]
  \centering
  \includegraphics[scale=.4]{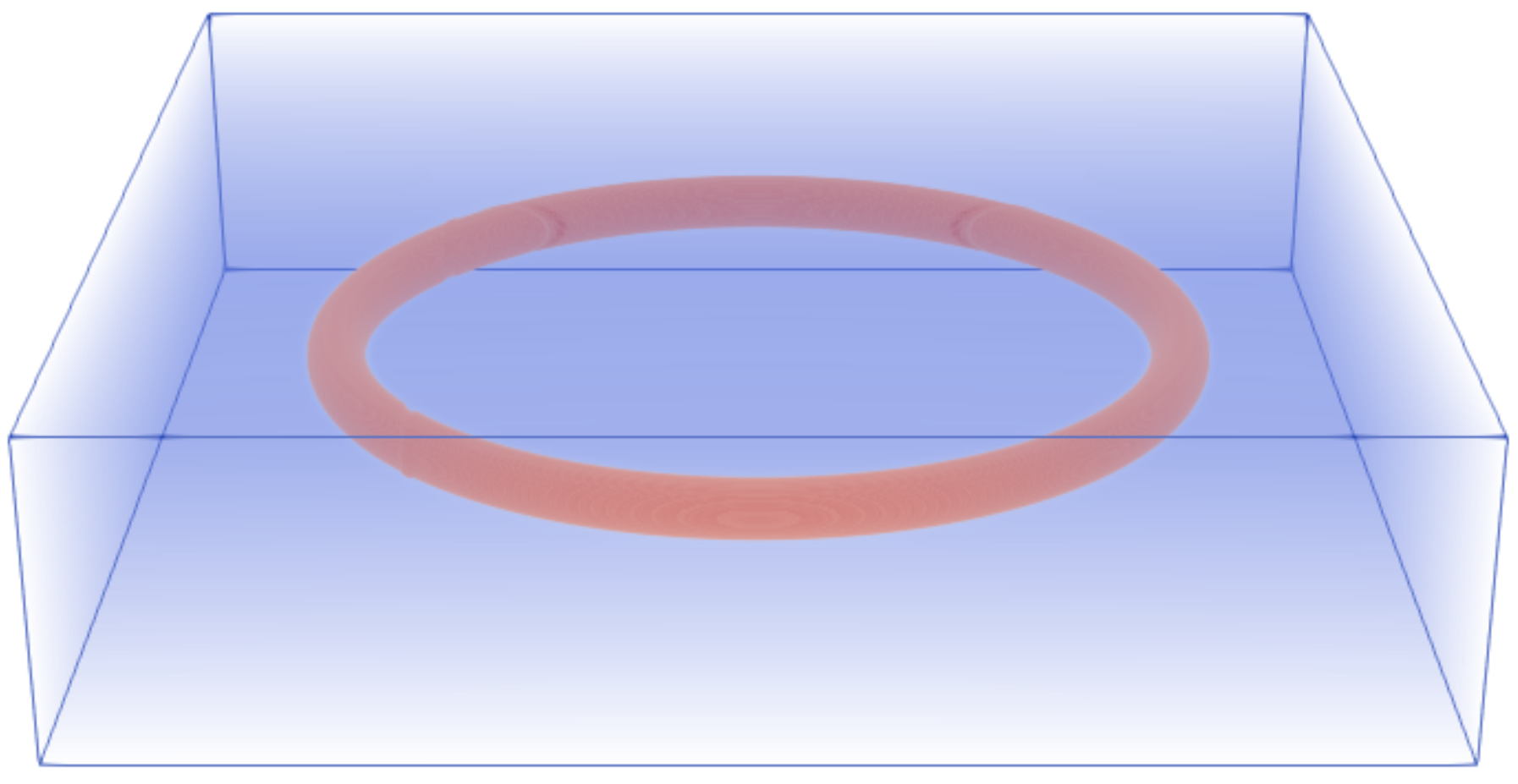}
  \hspace{.1in}
   \includegraphics[scale=.4]{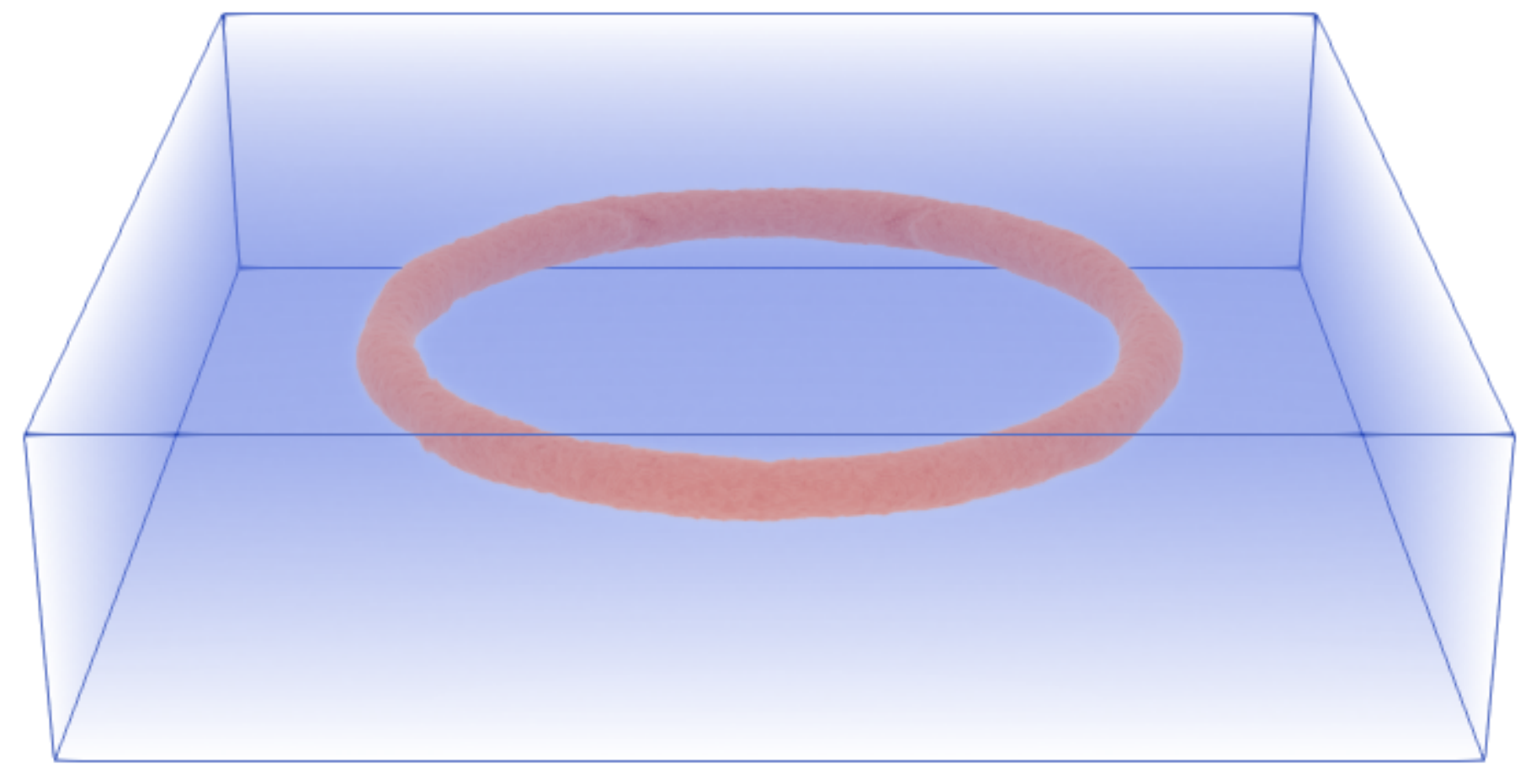}

   \includegraphics[scale=.4]{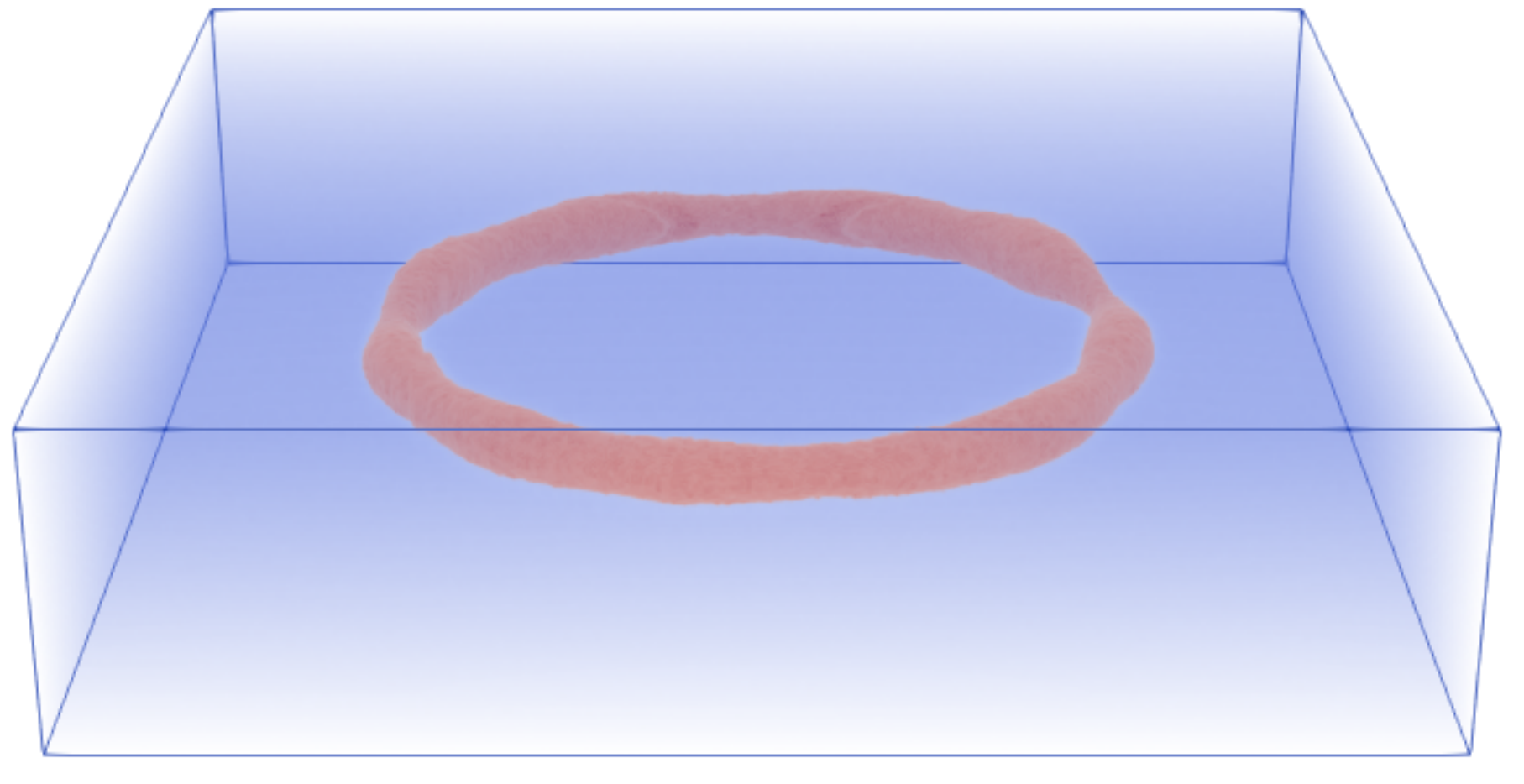}
   \hspace{.1in}
   \includegraphics[scale=.4]{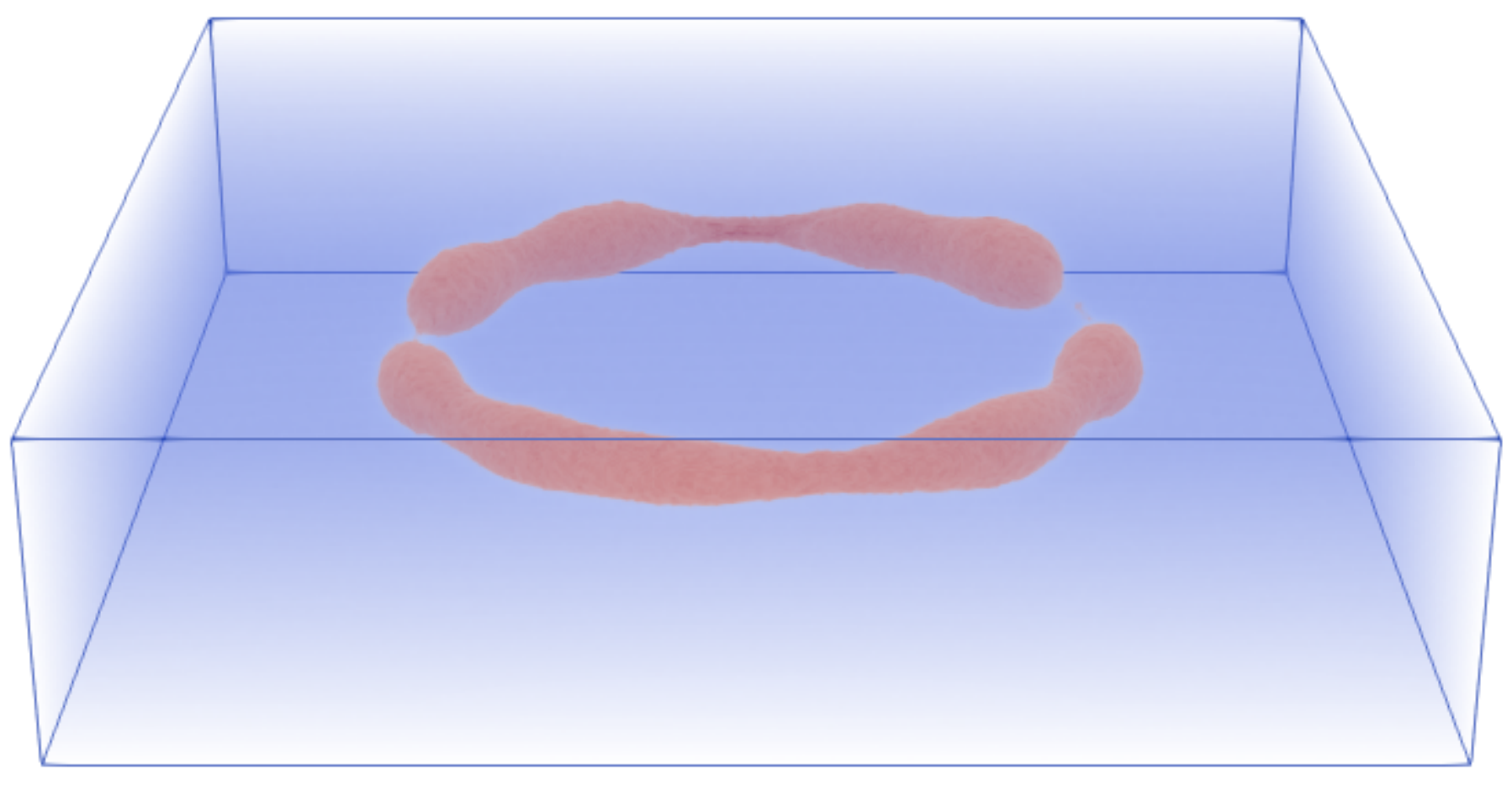}

   \includegraphics[scale=.4]{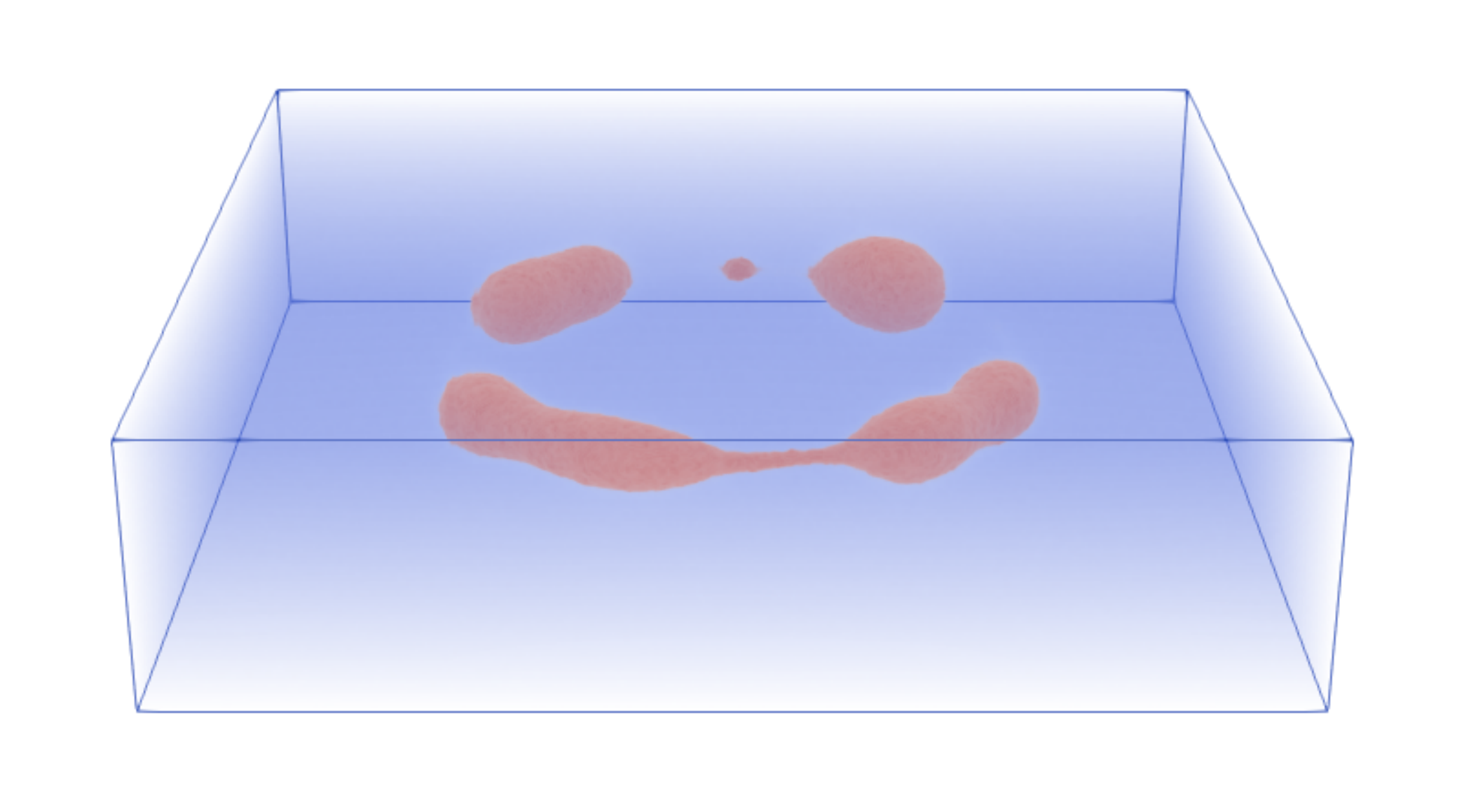}
   \hspace{.1in}
   \includegraphics[scale=.4]{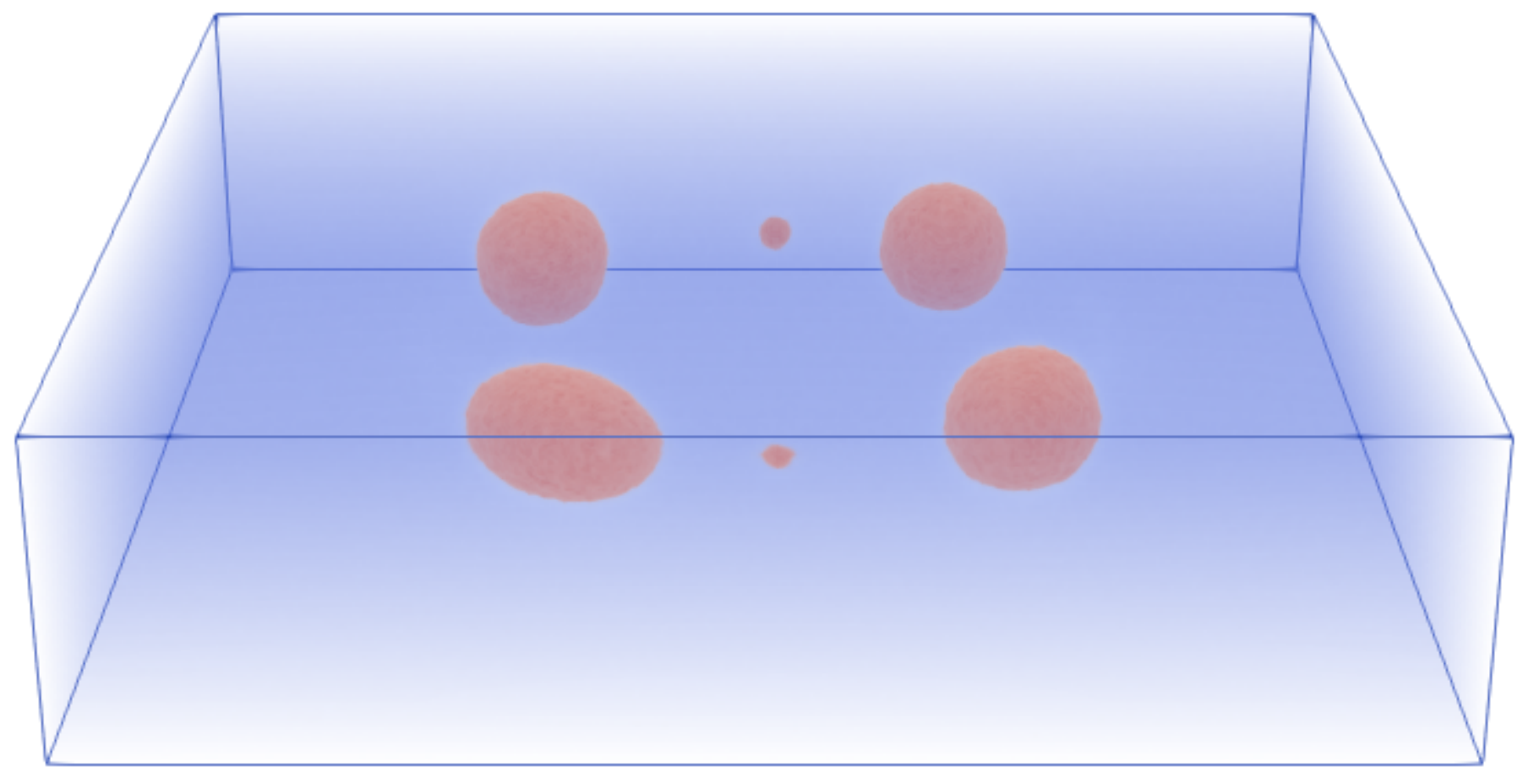}
   
  \caption{Rayleigh-Plateau instability on a torus; snapshots show initial data and simulation results at t = 4.0~ns, 6.0~ns, 8.0~ns, 10.0~ns, and 12.0~ns.}
  \label{fig:DonutShapshots}
\end{figure}

Finally, as an illustration of the capabilities of the algorithm we performed simulations showing the Rayleigh-Plateau instability on a torus.
In a 256.0~nm by 256.0~nm by 64.0~nm periodic system a torus with a center-line radius of 86.0~nm and a cylindrical radius of 6.0~nm was initialized.
The volumetric snapshots in Figure~\ref{fig:DonutShapshots} show the outer radius shrinking as the instability develops \cite{TorusShrink2011}, which is observed in macroscopic experiments \cite{TorusRayPlatPRL2009} and the appearance of satellite droplets, as predicted by theory \cite{TorusRayPlat2013}.
The impact of thermal fluctuations in this geometry remains an open question for future study.

\section*{Summary and conclusions}

The behavior of hydrodynamic instabilities play a crucial role in determining the dynamics of fluid 
systems. At the nanoscale the relative importance of various physical phenomena changes compared to macroscale systems. In particular thermal fluctuations can significantly influence behavior.
To assess the importance of thermal fluctuations in nanofluid systems a
 new dimensionless parameter, the stochastic Weber number, is required.
The results in this paper demonstrate thermal fluctuations can influence the growth and morphologies of nanostructures, specifically the pinching of liquid cylinders into droplets.

As discussed in the paper, this is not a new observation and our results are in good agreement with previous nanofluidic studies of the Rayleigh-Plateau instabilities using stochastic lubrication theory and molecular dynamics simulations.
The significance of the present work is in developing a numerical method for using fluctuating hydrodynamics effectively without requiring the lubrication approximation.
Fluctuating hydrodynamics offers several advantages compared with molecular dynamics.
Fluctuating hydrodynamics simulations allow us to set independently such physical parameters as surface tension and viscosity, which in molecular dynamics are indirectly linked to the intermolecular potentials.
Our earlier work demonstrates that fluctuating hydrodynamics calculations are also typically several orders of magnitude faster than molecular dynamics simulations.
Finally, in fluctuating hydrodynamics the modeling of complex fluids (e.g., ionic liquids, reactive mixtures) is straight-forward \cite{RTIL,donev2019electroneutral,kim2018reactiveFHD}.

The methodology developed here can be used to more fully quantify the role of diffusion \cite{DiffusionRayPlat2022}
and viscosity on the Rayleigh-Plateau instability.  To facilitate this type of study we are developing an implicit diffusion solver that will greatly improved the efficiency of the algorithm. The approach can also be extended to the case of fluids
with dissimilar properties such as density, viscosity and diffusivity.  Fluctuating hydrodynamics can also be used to investigate the behavior of thin films and model the dynamics of contact lines \cite{CapWaveThinFilm_2021,ThinFilmFluct_2021,CapWaveFluct2023}.
The overall approach can also be generalized to systems with more than two components including polymer mixtures, enabling the simulation of a wide range of multiphase phenomena at the nanoscale.

\FloatBarrier

\textbf{Acknowledgements} \Add{The authors thank Prof. Aleksander Donev for fruitful discussions.} This work was supported by the U.S. Department of Energy, Office of Science, Office of Advanced Scientific Computing Research, Applied Mathematics Program under contract No. DE-AC02-05CH11231. 
This material is based upon work supported by the U.S. Department of
Energy, Office of Science, Office of Advanced Scientific Computing Research, Department of
Energy Computational Science Graduate Fellowship under Award Number DE-SC0022158.
This research used resources of the National Energy Research Scientific Computing Center, a DOE Office of Science User Facility supported by the Office of Science of the U.S. Department of Energy under Contract No. DE-AC02-05CH11231.

\bigskip \hrule \bigskip

\section*{Supporting Information}

\subsection{Details of the discretization}

\begin{figure}
  \centering
  \includegraphics[scale=1.0]{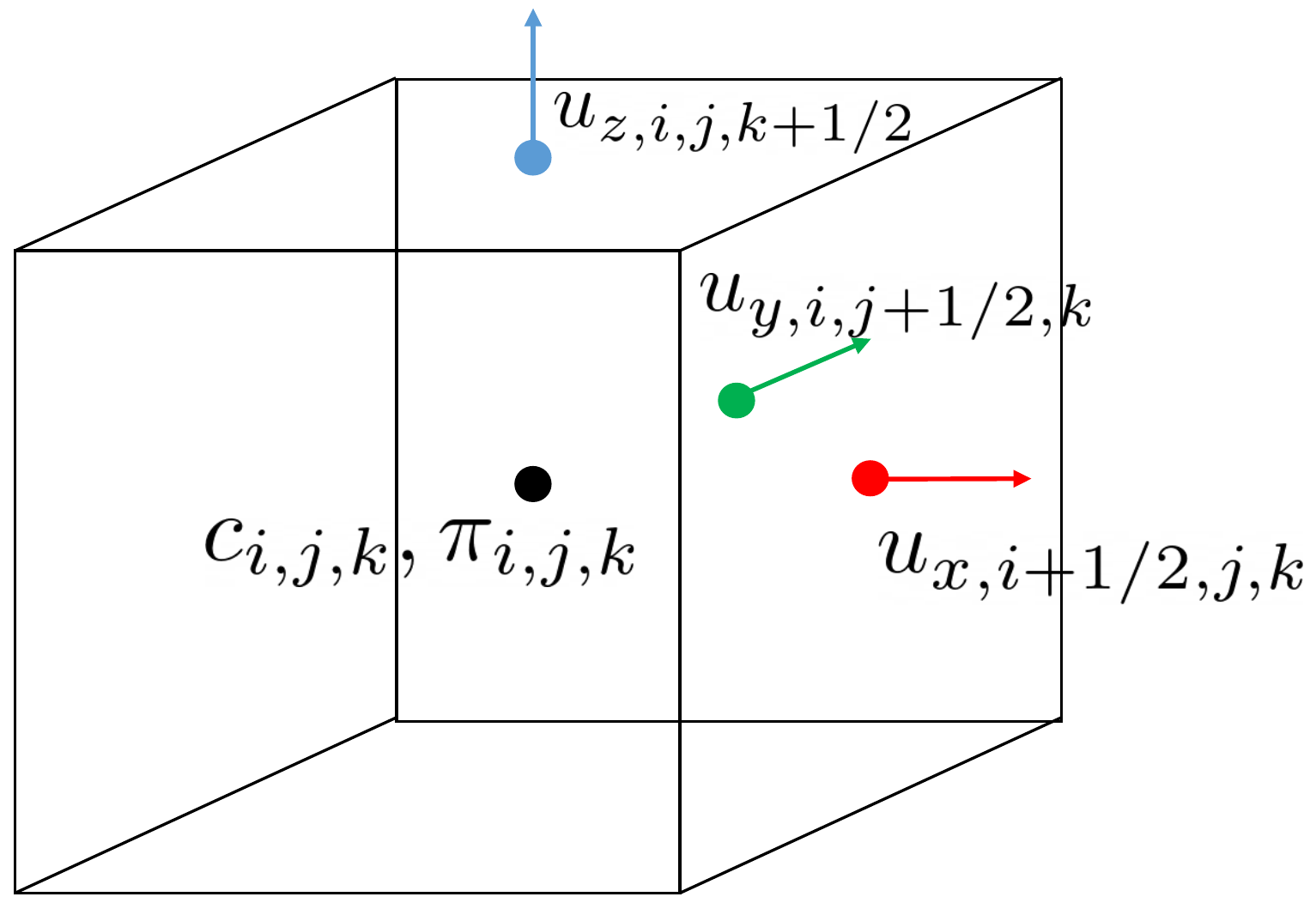}
  \caption{Sketch showing location where variables are defined in the staggered grid discretization.
  }
  \label{fig:mesh}
\end{figure}

The discretization uses a staggered grid representation on a uniform mesh with grid spacing $\Delta x$, $\Delta y$ and $\Delta z$ as depicted in Figure \ref{fig:mesh}.  Here the normal velocities in the $x$, $y$ and $z$ directions are given on faces denoted by $i+\half,j,k$, {$i,j+\half,k$} and $i,j,k+\half$, respectively.  Concentration and perturbational pressure are given on cell centers denoted by $i,j,k$.
The spatial discretizations of the convective terms in the concentration and momentum equation as well as the viscous stress are based on standard second-order stencils for derivatives and spatial averaging, as discussed in detail in~\cite{donev2014low, Donev_10}. Here we focus on discretization of the species diffusion and the reversible stress term.

\subsubsection{Species diffusion}

The divergence of the species flux is given by
\begin{equation}
(\nabla \cdot \SpeciesFlux )_{i,j,k} = \frac{\SpeciesFlux_{x,i+\half,j,k}-\SpeciesFlux_{x,i-\half,j,k}}{\Delta x}
+\frac{\SpeciesFlux_{y,i,j+\half,k}-\SpeciesFlux_{y,i,j-\half,k}}{\Delta y}
+\frac{\SpeciesFlux_{z,i,j,k+\half}-\SpeciesFlux_{z,i,j,k-\half}}
{\Delta z}
\label{eq:spec_div}
\end{equation}
The flux at an $x$ face is given by
\begin{align}
\overline{\SpeciesFlux}_{x,i+\half,j,k} = \rho D \left ( \left( 1 - 2 \chi  c (1-c) \right )\frac{\partial c}{\partial x} +  c(1-c) \kappa \frac{\partial  ( \nabla^2 c) }{\partial x} \right )_{i+\half,j,k} \;\;\; .
\end{align}
To evaluate this term we first approximate
\begin{align}
    (\nabla^2 c)_{i,j,k} \approx (\nabla_h^2 c)_{i,j,k} =
    \frac{c_{i-1,j,k} - 2 c_{i,j,k} + c_{i+1,j,k}}{\Delta x ^2}
+     \frac{c_{i,j-1,k} - 2 c_{i,j,k} + c_{i,j+1,k}}{\Delta y ^2}
   +   \frac{c_{i,j,k-1} - 2 c_{i,j,k} + c_{i,j,k+1}}{\Delta z ^2} 
\end{align}
We then approximate
\begin{align}
\overline{\SpeciesFlux}_{x,i+\half,j,k} \approx \rho D \left[ (1 - 2 \chi  c (1-c))_{i+\half,j,k} \frac{c_{i+1,j,k}-c_{i,j,k}}{\Delta x} +  ( c(1-c)  \kappa)_{i+\half,j,k} \frac{(\nabla_h^2 c)_{i+1,j,k}-(\nabla_h^2 c)_{i,j,k}}{\Delta x}  \right ]  
\end{align}
where the coefficients at $i+\half,j,k$ are evaluated by averaging from the two adjacent cells.
The other directions are treated analogously.

\subsubsection{Reversible stress}

The discretization of the reversible stress is somewhat more complicated because of the staggered representation of velocity. The divergence of the reversible stress in the $x$ momentum equation is
given by 
\begin{align}
(\nabla \cdot R)_x = 
\frac{\rho k_B T \kappa}{m} \left( \pderiv{R_{xx}}{x}+ \pderiv{R_{xy}}{y}+ \pderiv{R_{xz}}{z}\right )
\end{align}
where
\[
R_{xx} = \frac{1}{2} ( c_y^2+c_z^2-c_x^2 ) \;, \;\;\;\;R_{xy} = -c_x c_y \;,\;\;\;\mathrm{and} \;\;\;R_{xz} = -c_x c_z 
\]
with $c_\alpha = \partial c/\partial \alpha$. We need to discretely evaluate $(\nabla \cdot R)_x$ at the cell face where the $x$ velocity is defined.  Thus, we discretize
\[
(\nabla \cdot R)_{x,i+\half,j,k} = 
\frac{\rho k_B T \kappa}{m} \left[ 
\frac{R_{xx,i+1,j,k}-R_{xx,i,j,k}}{\Delta x}+
\frac{R_{xy,i+\half,j+\half,k}-R_{xy,i+\half,j-\half,k}}{\Delta y}+
\frac{R_{xx,i+\half,j,k+\half}-R_{xz,i+\half,j,k-\half}}{\Delta z}
\right]
\]
Note that $R_{xx}$ is defined at cell centers whereas $R_{xy}$ and $R_{xz}$ are defined on edges.

To define these terms we first define gradients of the concentration on nodes
\begin{align}
  ( G^n_x c)_{i+\half,j+\half,k+\half} &= \frac{1}{4\Delta x} \left(
  c_{i+1,j+1,k+1}+ c_{i+1,j+1,k}+c_{i+1,j,k+1}+c_{i+1,j,k}- c_{i,j+1,k+1}- c_{i,j+1,k}-c_{i,j,k+1}-c_{i,j,k}\right) \\
   ( G^n_y c)_{i+\half,j+\half,k+\half} &= \frac{1}{4\Delta y} \left(
  c_{i+1,j+1,k+1}+ c_{i+1,j+1,k}+c_{i,j+1,k+1}+c_{i,j+1,k}- 
   c_{i+1,j,k+1}- c_{i+1,j,k}-c_{i,j,k+1}-c_{i,j,k}\right) \\
   ( G^n_z c)_{i+\half,j+\half,k+\half} &= \frac{1}{4\Delta z} \left(
  c_{i+1,j+1,k+1}+ c_{i+1,j,k+1}+c_{i,j+1,k+1}+c_{i,j,k+1}- c_{i+1,j+1,k}- c_{i+1,j,k}-c_{i,j+1,k}-c_{i,j,k}\right) 
\end{align}
We can now define $R_{xx}$ at cell centers by averaging the gradients to cell centers.  More precisely, we approximate
\begin{align}
(\nabla c)_{i,j,k}
\approx \frac{1}{8} \bigl(
 ( &\mathbf{G}^n c)_{i+\half,j+\half,k+\half}+
  ( \mathbf{G}^n c)_{i+\half,j+\half,k-\half}+
   ( \mathbf{G}^n c)_{i+\half,j-\half,k+\half}+
    ( \mathbf{G}^n c)_{i+\half,j-\half,k-\half}+ \nonumber \\
     ( &\mathbf{G}^n c)_{i-\half,j+\half,k+\half}+
      ( \mathbf{G}^n c)_{i-\half,j+\half,k-\half}+
       ( \mathbf{G}^n c)_{i-\half,j-\half,k+\half}+
        ( \mathbf{G}^n c)_{i-\half,j-\half,k-\half} \bigr)
\end{align}
where $\mathbf{G}^n = (G^n_x,G^n_y,G^n_z)$.  Using these approximatation we can now define $R_{xx}$
at cell centers.

For the other terms we define
\[
R_{xy,i+\half,j+\half,k} \approx \frac{1}{2} \left (
 ({G}^n_x c)_{i+\half,j+\half,k+\half}\times
  ( {G}^n_y c)_{i+\half,j+\half,k+\half}+
   ( {G}^n_x c)_{i+\half,j+\half,k-\half}\times
    ( {G}^n_y c)_{i+\half,j+\half,k-\half} \right)
\]
and
\[
R_{xz,i+\half,j,k+\half} \approx \frac{1}{2} \left (
 ({G}^n_x c)_{i+\half,j+\half,k+\half}\times
  ( {G}^n_z c)_{i+\half,j+\half,k+\half}+
   ( {G}^n_x c)_{i+\half,j-\half,k+\half}\times
    ( {G}^n_z c)_{i+\half,j-\half,k+\half} \right)
\]

\subsubsection{Discretization of noise}

The white noise terms, $\mathcal{Z}$ and $\mathcal{W}$, in the equations are used to compute an additional stochastic fluxes that represents thermal fluctuations.  These terms cannot be evaluated pointwise in either space or time.  In the discretization these white noise terms are represented in terms of a spatio-temporal average of a time interval of $\Delta t$ and a spatial region of size $\Delta x \times \Delta y \times \Delta z$.  In this integrated form, $\mathcal{Z}$ can be modeled as vector independent Gaussian random variables $Z$ with mean $0$ and variance 
\[
\sigma^2 = \frac{1}{\Delta t \, \Delta x \, \Delta y \, \Delta z} \;\;\;.
\]
Similarly, $\mathcal{W}$ can be modeled as a matrix of independent Gaussian random variables with the same $\sigma^2$.

\begin{figure}
  \centering
  \includegraphics[scale=1.0]{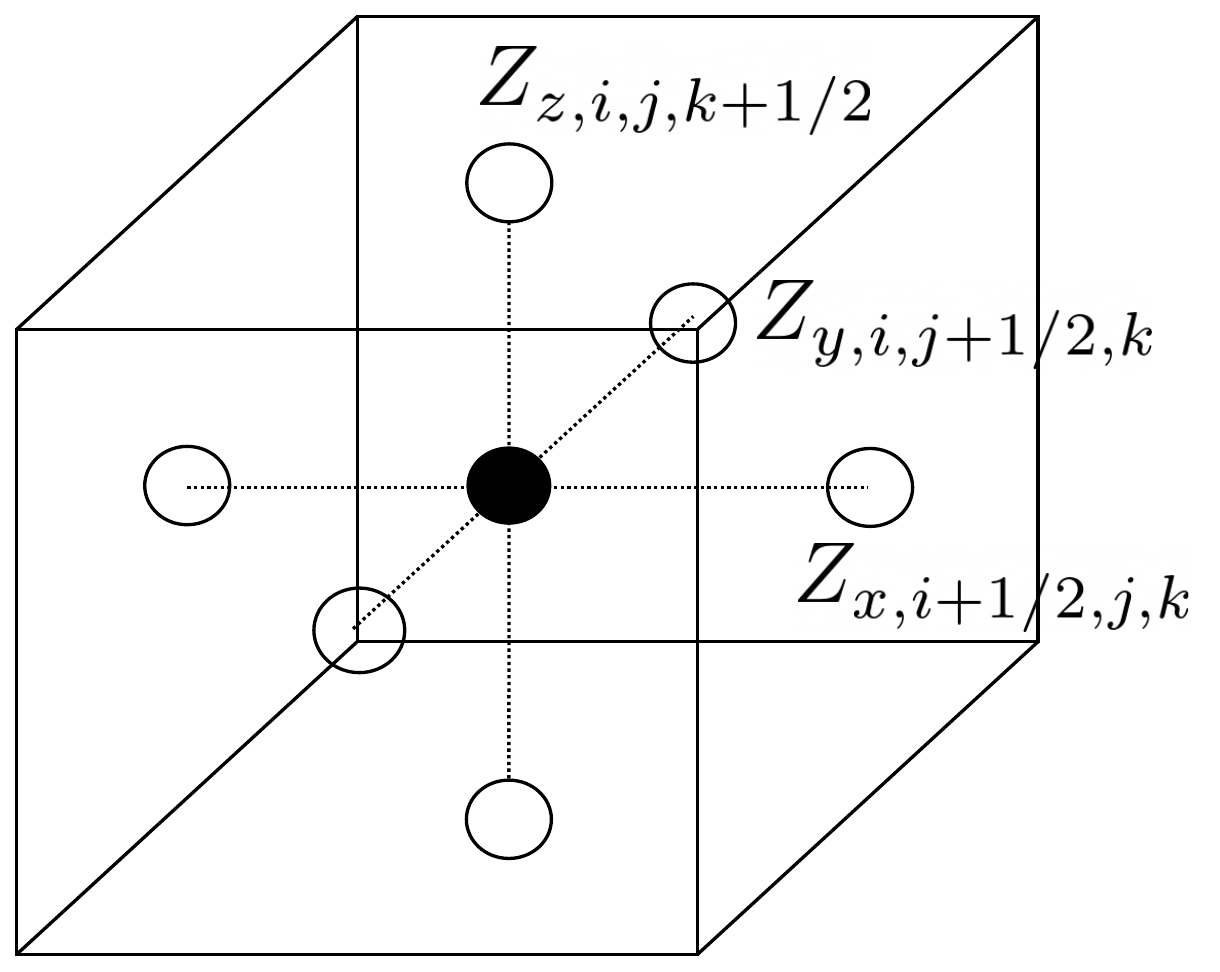}
  \caption{Sketch showing location where stochastic species fluxes are computed.
  }
  \label{fig:spec_noise}
\end{figure}

The components of $Z$ are generated at faces as illustrated in Figure \ref{fig:spec_noise}; i.e., we define
\[
Z_{x,i+\half,j,k} \;\;\;Z_{x,i,j+\half,k} \;\;\; \mathrm{and} \;\;\;Z_{z,i,j,k+\half} 
\]

The stochastic species fluxes are then approximated by, for example,
\begin{align}
\widetilde{\SpeciesFlux}_x = \sqrt {2\rho m D c_{i+\half,j,k} (1-c_{i+\half,j,k})} ~ Z_{x,i+\half,j,k}.
\end{align}
The stochastic fluxes {can} then be added to the deterministic species fluxes prior to the evaluation of Eq. (\ref{eq:spec_div}).

\begin{figure}
  \centering
  \includegraphics[scale=.6]{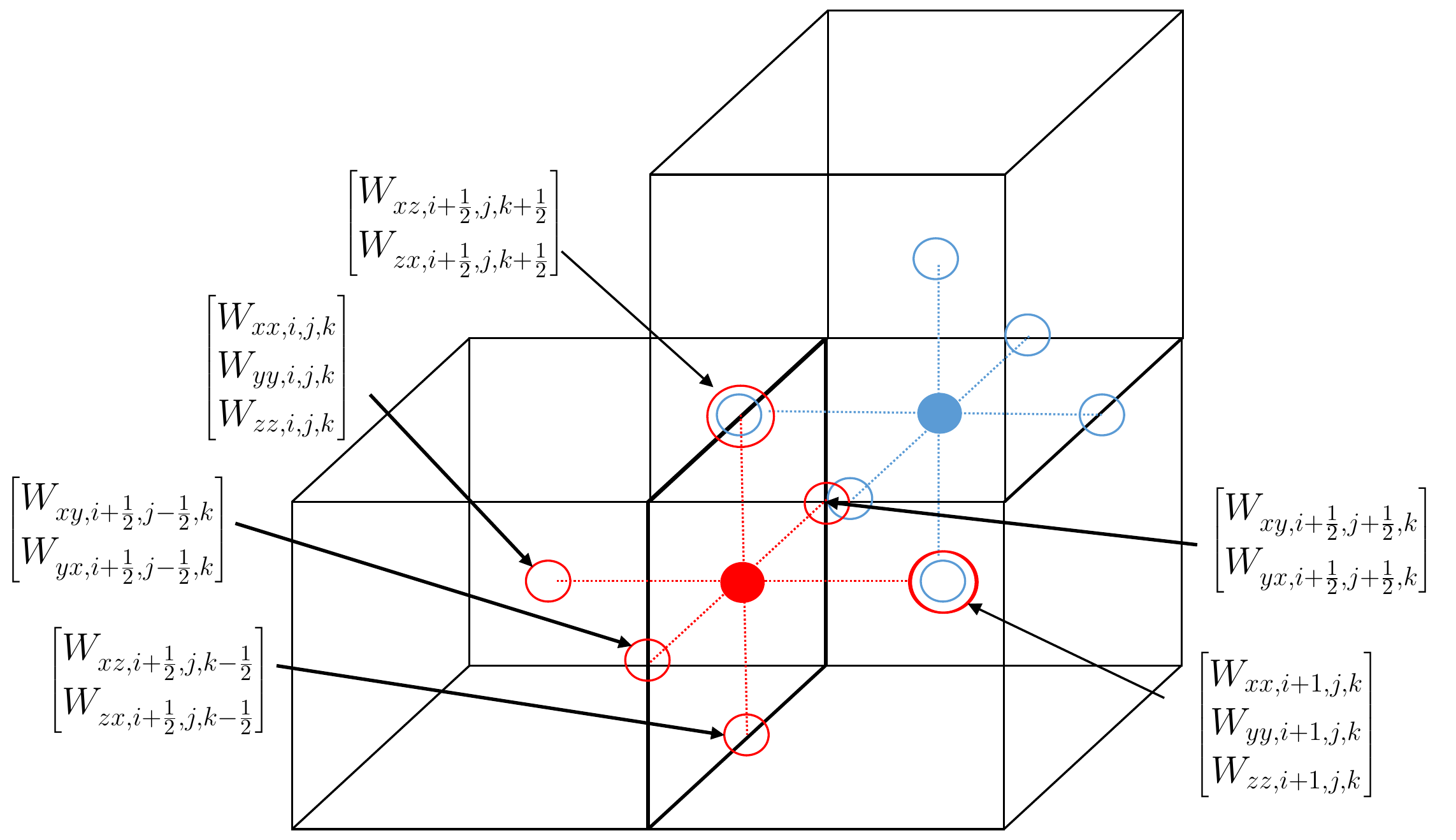}
  \caption{Sketch showing location where different components of the stochastic stress tensor are computed. Here the red points are used to compute $(\nabla \cdot (W + W^T))_x$ at face $i+\half,j,k$.  The cyan points are used to compute $(\nabla \cdot (W + W^T))_z$ at face $i+1,j,k+\half$.
  }
  \label{fig:tensor_noise}
\end{figure}

Because of the staggered grid representation of velocities, the terms in $W$ representing the Gaussian random field are associated with a number of different locations, as shown in Figure \ref{fig:tensor_noise}.

In particular, the diagonal entries, $W_{xx}, W_{yy} $ and $W_{zz}$ are computed at cell centers denoted by $i,j,k$.  The off diagonal terms are defined at the centers of edges.  Specifically, $W_{xy}$ and $W_{yx}$ are computed at the centers of the edge going from node $(i+\half,j+\half,k-\half)$ to node $(i+\half,j+\half,k+\half)$, which are denoted as $i+\half,j+\half,k$.  We define
\[
W_{xz,i+\half,j,k+\half} \;\;\; W_{zx,i+\half,j,k+\half} \;\;\;
W_{yz,i,j+\half,k+\half} \;\;\; \mathrm{and} \;\;\; {W_{zy,i,j+\half,k+\half}} 
\]
analogously.
The $x$ component of $\nabla \cdot (W + W^T)$ is then discretized as
\begin{align*}
\left[\nabla \cdot (W + W^T)\right ]_{x,i+\half,j,k}&=
2\frac{W_{xx,i+1,j,k}-W_{xx,i,j,k}}{\Delta x} \\
&+
\frac{W_{xy,i+\half,j+\half,k}+W_{yx,i+\half,j+\half,k}
-W_{xy,i+\half,j-\half,k}+W_{yx,i+\half,j-\half,k}}{\Delta y} \\
&+
\frac{W_{xz,i+\half,j,k+\half}+W_{zx,i+\half,j,k+\half}
-W_{xz,i+\half,j,k-\half}+W_{zx,i+\half,j,k-\half}}{\Delta x}
\end{align*}
The other components are treated analogously.

\subsection{Measuring the interface thickness}

\begin{figure}
  \centering
  \includegraphics[scale=1.0]{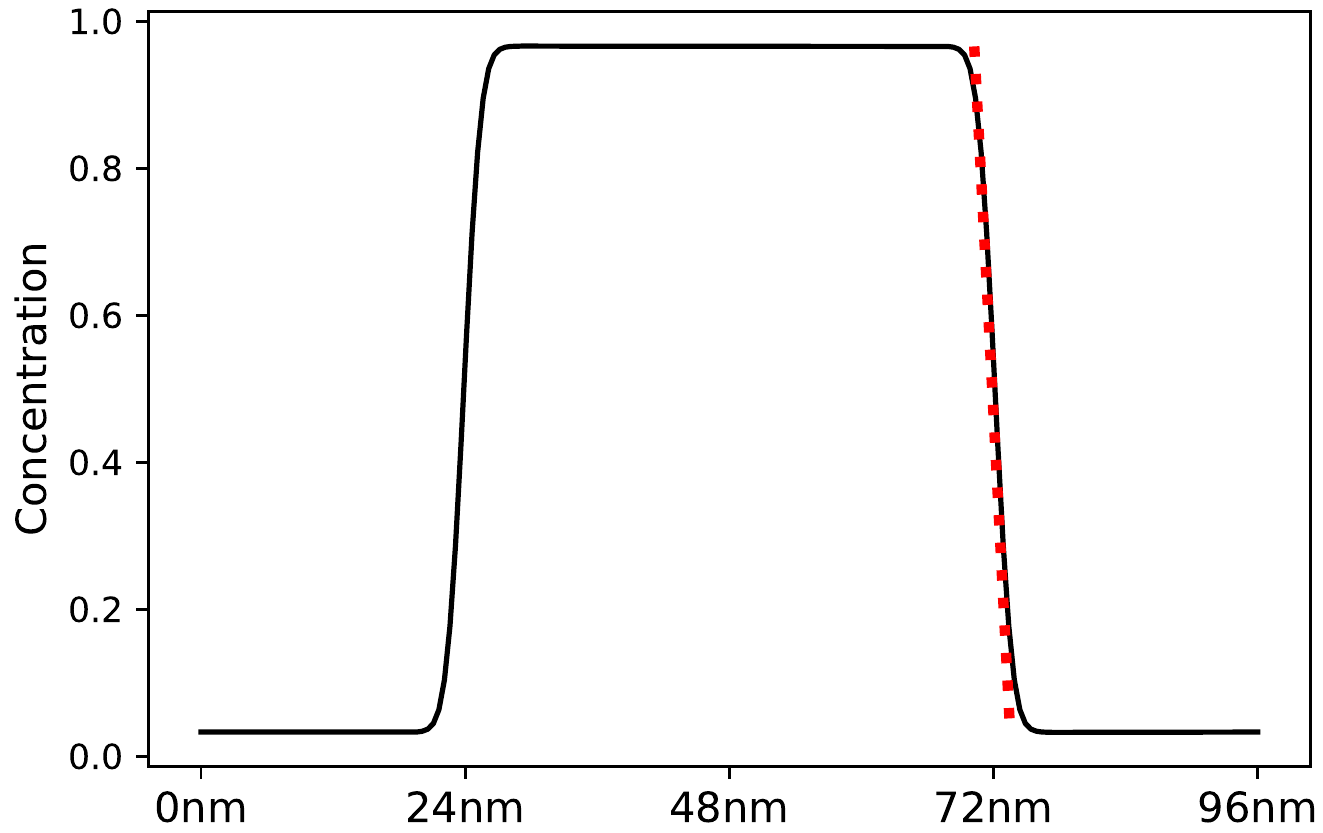}
  \caption{The interface thickness is computed from the tangent line at the center of the interface extrapolated to the inner and outer values of concentration.
  }
  \label{fig:ThicknessExample}
\end{figure}

\begin{figure}
  \centering
  \includegraphics[scale=0.6]{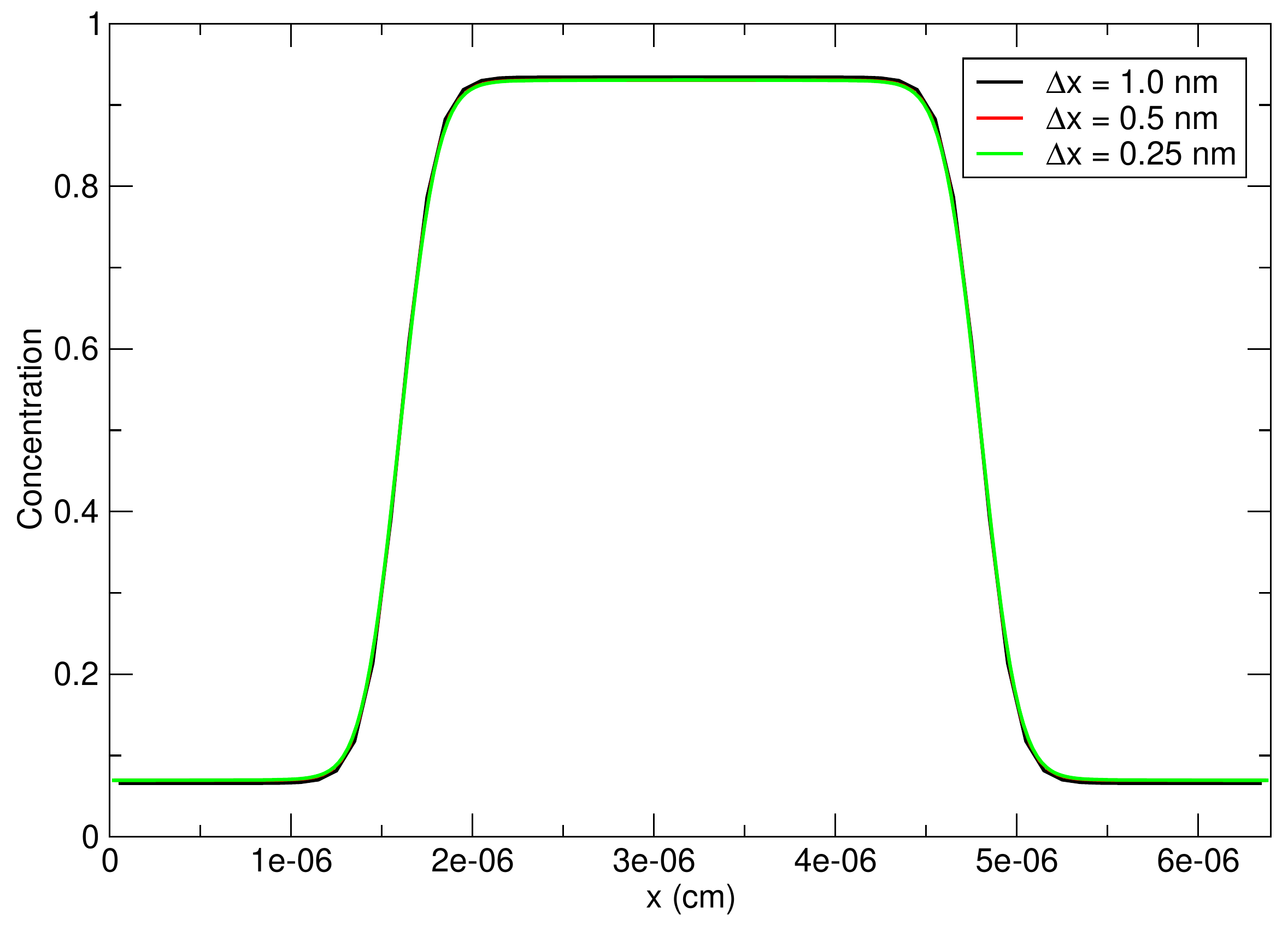}
  \caption{\Add{Numerical results showing slice through slab at three different resolutions for $\chi = 3.0$.}}
  \label{fig:slice}
\end{figure}

First, a slab initiated inside and outside with concentrations $c_{int}$ and $c_{ext}$ is run to equilibrium (see Fig.~\ref{fig:ThicknessExample}). We approximate the interface thickness using the line tangent to the inflection point along the interface extrapolated to these concentrations. Specifically, given the slope $S$ of this tangent line,
we define $\ell_\mathrm{s}$ to be the distance required for a line with that slope to change from the interior to the exterior concentration; that is,
\[
\ell_\mathrm{s }S = |c_{ext}-c_{int}| \;\;\;. \]
The inflection point is approximated as the maximum of $|\nabla c|$, 
which is computed via second order centered differencing in the interior.
\Add{We can use the resulting approximated slope to estimate the interface thickness as
\[
{\ell_\mathrm{s}} \approx \left|~ (c_{ext}-c_{int}) \left/ \left(\frac{c_{j+1}-c_{j-1}}{x_{j+1}-x_{j-1}}\right)\right|\right.,
\]
with the $j^\mathrm{th}$ grid point having the maximum $|\nabla c|$ in the discrete data. 

We note that this computation of  ${\ell_\mathrm{s}}$ is sensitive to numerical resolution.  In Fig.~\ref{fig:slice} we show the computed profiles for $\chi = 3.0$ at three different resolutions, illustrating the resolution of the methodology.  We estimate the thickness using the procedure discussed above to obtain $\ell_\mathrm{s} = 4.34 \, \mathrm{nm}, \; 4.20 \, \mathrm{nm},\; \mathrm{and} \; 4.17 \, \mathrm{nm}$ corresponding to $\Delta x = 1.0 \,\mathrm{nm},\; 0.5\, \mathrm{nm}, \; \mathrm{and}\; 0.25\, \mathrm{nm}$, respectively where the theoretical result is $4.16 \,\mathrm{nm}$. This demonstrates second-order convergence of the computed thickness. 

\subsection{Measuring disk radius for Laplace pressure validation test}

The computation of the interfacial tension from the Laplace pressure depends on how the radius of the disk is measured.  Here we measure the distance from the center of the disk to the point at which $c = 0.5$ based on linear interpolation of the simulation data.  The data in the graph is taken from a horizontal slice through the center of the disk.  Tests of other angles (e.g., diagonal slice) at this resolution ($\Delta x = 1.0$ nm) showed roughly a 1\% variation.  Additional numerical tests at finer resolution showed that simulations converged toward the theoretical result with increasing resolution.

}

\subsection{Calculation of cylinder radii}

\begin{figure}
  \centering
  \includegraphics[scale=.4]{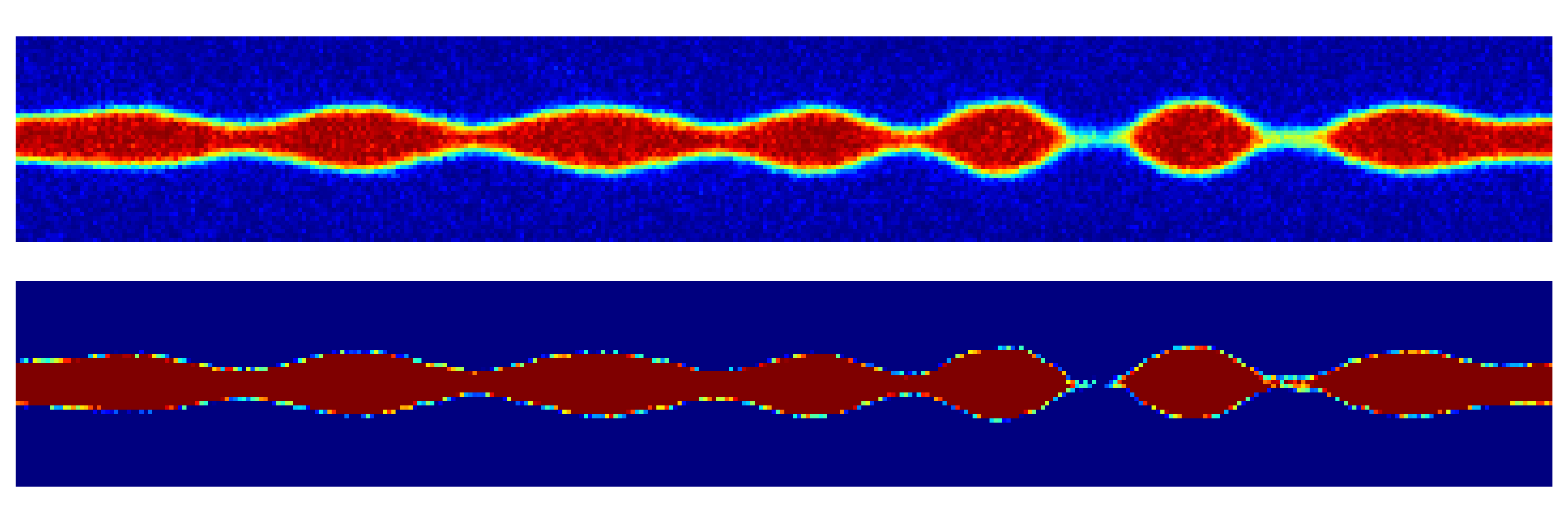}
  \caption{Visual comparison of the raw data (top) and the filtered data (bottom) of the cylinder at pinch time.
  }
  \label{fig:RawVsFiltered}
\end{figure}

Because we use a diffuse interface model the raw data for concentration is filtered for the purpose of calculating the cylinder radii. Specifically, the filtered concentration is
\begin{align}
    \tilde{c} = \max \left( \min \left( \frac{ c + \Delta c_f - 1/2}{2\Delta c_f} , 1 \right), 0\right)
\end{align}
This filter modifies the thickness of an interface, specifically $\tilde{\ell}_\mathrm{s} = 2 \Delta c_f \ell_\mathrm{s}$; for our calculations we chose $\Delta c_f = 0.1$ so the filter reduces the interface thickness by a factor of five. 
The radius for a slice of cells is then computed as 
\begin{align}
    R(z,t)=\sqrt{\frac{1}{\pi}(\Delta x)(\Delta y)\sum_{i=1}^{N_x}\sum_{j=1}^{N_y}
    \tilde{c}_{i,j,k}^n},
\end{align} 
where $z = k \Delta z$ and $t = n \Delta t$. A side-by-side comparison of the raw concentration data and the filtered concentration data for a stochastic run (Oh = 0.5) at the approximated pinch time is shown in  Fig.~\ref{fig:RawVsFiltered}.

\FloatBarrier

\subsection{Fitting power laws to the minimum radius versus time data}

To compute power-law fits to the minimum radius versus time, we first identify the pinch time, $t^\star$ for each simulation using the calculation of $R(z,t)$ described above.  We then recast the data in terms of $t^\star -t$ for each simulation.  The resulting data are then averaged to obtain an ensemble for each set of conditions.  

For each case we then identify an interval over which to perform the fit.  In particular, we exclude the region near where $R \rightarrow 0$. This avoids sensitivity of the fit to to criterion used for identifying pinch time and the noisiness of the data at subgrid scales, particularly for the stochastic cases.  The resulting data can then be fit using linear regression.

\FloatBarrier

\subsection{Computing spectra using FFT}

For the computation of the spectra in Figures 7 and 8, we have defined
\[
\hat{R}(k,t) = \frac{L}{R_0} \left ( \frac{1}{L} \int_0^L {R(z,t)} e^{-i k z} dz \right) \;\;\;.
\]
where the term in parenthesis is computed using a standard FFT routine.
The factor of $L/R_0$ is introduced so that the spectra plotted in Figs. 7 and 8 use the same normalization as used by Zhao {\it et al.} \cite{RayPlatFHD}.  \Add{Note that in Figure 3 the standard FFT (i.e., without the prefactor) is used to obtain $\hat{h}$.}

\FloatBarrier

\FloatBarrier

\FloatBarrier

\FloatBarrier

\FloatBarrier

\FloatBarrier

\FloatBarrier

\bigskip \hrule \bigskip

\bibliography{RayPlat}

\end{document}